\documentclass[12pt]{article}

\newcommand{\rem}[1]{}
\newcommand{\remv}[1]{}

\newcommand{\commentstarts}{\begin{centering}
\hspace{-1pt}\vrule\vrule
\begin{minipage}[t]{0.03\linewidth}
\hspace{0.025\linewidth}
\end{minipage}
\begin{minipage}[t]{0.95\linewidth}}
\newcommand{\commentends}{\end{minipage}
\end{centering}
\vspace{7pt}
}

\usepackage{graphicx}
\usepackage{alltt}
\usepackage{amsmath}
\usepackage{amssymb}
\usepackage{hyperref}

\def\amshow{0}

\ifodd\amshow

\fi

\begin{document} 

\newcounter{Conjectures}
\setcounter{Conjectures}{0}
\newcounter{Lemmas}
\setcounter{Lemmas}{0}
\newcounter{Theorems}
\setcounter{Theorems}{0}

\newcommand{\ashriek}{\mbox{\small !`}}

\begin{titlepage}
\begin{flushright}

\end{flushright}

\begin{center}
{\Large\bf $ $ \\ $ $ \\
On the construction of integrated vertex in the pure spinor formalism
in curved background
}\\
\bigskip\bigskip\bigskip
{\large Andrei Mikhailov${}^{\dag}$}
\\
\bigskip\bigskip
{\it Instituto de F\'{i}sica Te\'orica, Universidade Estadual Paulista\\
R. Dr. Bento Teobaldo Ferraz 271, 
Bloco II -- Barra Funda\\
CEP:01140-070 -- S\~{a}o Paulo, Brasil\\
}

\vskip 1cm
\end{center}

\begin{abstract}
We have previously described a way of describing the relation between  
unintegrated and integrated vertex operators in $AdS_5\times S^5$ which uses the
interpretation of the BRST cohomology as a Lie algebra cohomology and 
integrability properties of the AdS background. Here we clarify some details
of that description, and develop a similar approach for an arbitrary
curved background with nondegenerate RR bispinor. For an arbitrary curved background, the sigma-model
is not integrable. However, we argue that a similar construction still works
using an infinite-dimensional Lie algebroid.
\end{abstract}

\vspace{30pt}

\small{ 
\begin{tabular}{rl}
MSC classes: & {\tt 83E50 (Primary) 17B55, 16S37 (Secondary)}  \\  
& \\  
Keywords: & superstring theory, quadratic algebras, \\  
& pure spinor formalism, vertex operators \\  
\end{tabular}
}

\vfill
{\renewcommand{\arraystretch}{0.8}%
\begin{tabular}{rl}
${}^\dag\!\!\!\!$ 
& 
\footnotesize{on leave from Institute for Theoretical and 
Experimental Physics,}
\\    
&
\footnotesize{ul. Bol. Cheremushkinskaya, 25, 
Moscow 117259, Russia}
\\
\end{tabular}
}

\end{titlepage}

\tableofcontents
\section{Introduction}
The construction of the worldsheet sigma-model for the Type II superstring
in the pure spinor formalism is a fundamental problem. It was more or less
solved in \cite{Berkovits:2001ue}. However, we feel that some better understanding is possible.
First of all, the sigma-model suggested in \cite{Berkovits:2001ue} is technically very special,
and it is not clear why this is the most general solution. In particular,
the formulation depends crucially on a special choice of fields. Indeed, the 
theory is not invariant under field redefinitions mixing matter fields with
ghosts. It would be desirable to have an axiomatic formulation of the 
sigma-model. Something along these lines:
\begin{itemize}
\item A sigma-model with two nilpotent symmetries, $Q_L$ and $Q_R$, such that
the current of $Q_L$ is holomorphic and the current of $Q_R$ is antiholomorphic,
and there are symmetries $U(1)_L$ and $U(1)_R$, such that $Q_L$ and $Q_R$ are
appropriately charged under them.
\end{itemize}
However, we feel that this is not enough; the axiomatics sketched above is 
probably too weak, although it is enough to correctly describe small
deformations of the flat space. The correct axiomatics should somehow encode
the singularity of the pure spinor cone. 

Also, we believe that the worldsheet sigma-model should be formulated as a 
problem in cohomological perturbation theory. A small neighborhood of each 
point in space-time can be approximated by flat space:
\begin{equation}\label{FlatSpaceSigmaModel}
   S = \int d\tau^+ d\tau^- \left(
      \partial_+X^{\mu}\partial_-X^{\mu} + p_+\partial_-\theta_L 
      + p_-\partial_+\theta_R + w_+\partial_-\lambda_L + w_-\partial_+\lambda_R
   \right)
\end{equation}
with BRST symmetries:
\begin{align}
   Q_L \;&=\; \lambda_L^{\alpha}\left(
      {\partial\over\partial\theta_L^{\alpha}} +
      \Gamma^m_{\alpha\beta}\theta_L^{\beta}{\partial\over\partial x^m}
   \right) + (\ldots){\partial\over\partial w_+}
   \\    
   Q_R \;&=\; \lambda_R^{\hat{\alpha}}\left(
      {\partial\over\partial\theta_R^{\hat{\alpha}}} +
      \Gamma^m_{\hat{\alpha}\hat{\beta}}\theta_R^{\hat{\beta}}{\partial\over\partial x^m}
   \right) + (\ldots){\partial\over\partial w_-}
\end{align}
Then we say that a general background is obtained by the deformation
of the action accompanied by some deformation\footnote{As we have shown in \cite{Mikhailov:2012id}, the
very leading effect will be actually the deformation of $Q_L$ and $Q_R$ leaving
$S$ undeformed; this corresponds to the linear dilaton.} of $Q_L$ and $Q_R$. The 
infinitesimal deformations at the linearized level are well known to correspond 
to the linearized SUGRA waves. They were classified in \cite{Mikhailov:2014qka}. However, it was 
shown also in \cite{Mikhailov:2014qka} that there is a potential obstacle to extending the 
deformations beyond the linearized level. The obstacle is a nonzero cohomology 
group, namely the ghost number three vertex operators. Without doubt, the 
obstacle actually vanishes (there is a nonzero cohomology group, but the actual 
class vanishes). This, however, is not well understood. As we explained in \cite{Mikhailov:2014qka}, 
one way to prove the vanishing of the obstacle is to consider the action of the 
$b$-ghost in cohomology. The formalism that would allow to do such calculation
has not yet been fully developed. The definition of the $b$-ghost requires 
including the non-minimal fields which makes the lack of axiomatic formulation 
even more acute. And the $b$-ghost is nonpolynomial, opening the possibility 
that at some order the deformed action will also become non-polynomial\footnote{We have no doubt that this does
not happen, it is just that we don't know how to see this using the 
cohomological perturbation theory}. Such questions should be addressed together with the problem of axiomatic 
formulation of the worldsheet theory.

When we study the pure spinor formalism as a cohomological perturbation 
theory, one important technical aspect is the relation between integrated
and unintegrated vertex operators. The deformation of the action is described
by {\em integrated} vertices:
\begin{equation}
   S \rightarrow S + \int d\tau^+ d\tau^- \;U
\end{equation}
It is very important, that such deformations are in one-to-one correspondence
with the {\em unintegrated} vertices, which correspond to the cohomology of
$Q_L + Q_R$. One of the goals of this paper is to better understand the 
correspondence between integrated and unintegrated vertices.

In \cite{Mikhailov:2012uh,Mikhailov:2013vja,Chandia:2013kja} we have studied the relation between the pure spinor cohomology in
$AdS_5\times S^5$ and the Lie algebra cohomology, and argued that it is useful for 
understanding the relation between the integrated and unintegrated vertices. 
The pure spinor cohomology is the cohomology of the operator $Q_{\rm BRST}$ acting
on the space of functions $F(g,\lambda_L,\lambda_R)$:
\begin{equation}
   (Q_{\rm BRST}F)(g,\lambda_L,\lambda_R)=\left(
      \lambda_L^{\alpha}L_{\alpha} 
      +\lambda_R^{\hat{\alpha}}L_{\hat{\alpha}}
   \right) F(g,\lambda_L\lambda_R)
\end{equation}
Here $g\in G=PSU(2,2|4)$ and $L_{\alpha}$, $L_{\hat{\alpha}}$ are left shifts by some generators of
${\bf psu}(2,2|4)$. We introduced an infinite-dimensional Lie superalgebra ${\cal L}_{\rm tot}$, and
shown that the cohomology of $Q_{\rm BRST}$ is equivalent to some cohomology of ${\cal L}_{\rm tot}$.
Unintegrated vertices of the physical states correspond to the elements of the
second cohomology group. Moreover, there is a Lax pair $J_+$, $J_-$ taking values 
in ${\cal L}_{\rm tot}$. Given a cohomology class represented by a cocycle 
$\psi:\;\Lambda^2{\cal L}_{\rm tot}\; \to \;\mbox{Fun}(G)$, the corresponding integrated vertex is $\psi(J_+,J_-)$.
This construction uses special properties of $AdS_5\times S^5$.

\vspace{7pt}

Here we will describe a similar construction for an arbitrary curved spacetime
with the nondegenerate Ramond-Ramond field strength\footnote{Some elements of
our construction become degenerate if the Ramond-Ramond field strength is
zero. We do use the inverse of the RR bispinor $P^{\alpha\hat{\alpha}}$ in Section \ref{sec:Currents}. See the discussion of the flat space limit in \cite{Chandia:2013kja}.}.
Instead of the superalgebra ${\cal L}_{\rm tot}$ we will use some super Lie algebroid. We will
conjecture that the cohomology of this algebroid  is equal  to the BRST 
cohomology, {\it i.e.} unintegrated vertex operators. Moreover, there seems
to be an analogue of a Lax pair, which allows to construct integrated vertices. 
However, this Lax pair takes values in the sections of a Lie algebroid (instead
of a fixed Lie algebra), and presumably does not lead to integrability.

Better understanding of the integrated vertices could also help to explain
the consistency of the higher orders of the deformation of the action (\ref{FlatSpaceSigmaModel}).
Superficially, this problem looks similar to the PBW theorem of 
quadratic-linear algebras which (coincidentally?) is also useful in the 
construction of the integrated vertex. 

In eleven dimensional SUGRA, the analogous problem is the membrane worldsheet 
theory \cite{Berkovits:2002uc}. However, it appears more difficult than string worldsheet theory. 
But unintegrated vertices are more or less understood. Constructing integrated 
vertex operators is very close to understanding the worldsheet theory. Maybe 
some methods which we are developing here could be useful.

\paragraph     {Plan of the paper}
\begin{itemize}
   \item In Section \ref{sec:AdS} we give a streamlined review of \cite{Mikhailov:2012uh,Mikhailov:2013vja,Chandia:2013kja}, also simplifying some
      of the proofs in those references
   \item Section \ref{sec:GeneralCurved} develops a different point of view on the formalism of \cite{Berkovits:2001ue};
      our approach emphasizes the similarity between the constraints of the 
      Type IIB SUGRA and the constraints of the supersymmetric Yang-Mills
      theory
   \item In Section \ref{sec:Currents} we study the worldsheet currents. We construct an object
      resembling the Lax pair of the AdS theory, but using an algebroid instead
      of a Lie algebra. We conjecture that the cohomology of this algebroid
      corresponds to integrated vertex operators
\end{itemize}

\section{Brief review of the case of $AdS_5\times S^5$}\label{sec:AdS}
Here we will briefly review the relation between the unintegrated and integrated
vertices described in \cite{Mikhailov:2012uh,Mikhailov:2013vja,Chandia:2013kja}. Both types of vertices are obtained from the same
relative cocycle of some infinite-dimensional Lie superalgebra which we call 
${\cal L}_{\rm tot}$. In this sense, the relative Lie algebra cohomology of ${\cal L}_{\rm tot}$ provides the
unified description of both types of vertex operators.

\subsection{Definition of ${\cal L}_{\rm tot}$ and the PBW theorem}\label{sec:PBWforADS}
The infinite-dimensional superalgebra ${\cal L}_{\rm tot}$ is defined in \cite{Mikhailov:2013vja} by 
``gluing together'' two copies of the Yang-Mills algebra which we call ${\cal L}_L$ and 
${\cal L}_R$, in the following way. The ${\cal L}_L$ is generated by letters $\nabla^L_{\alpha}$, and ${\cal L}_R$ is 
generated by $\nabla^R_{\hat{\alpha}}$, satisfying the super-Yang-Mills constraints:
\begin{equation}\label{BasicRelations}
   \{\nabla_{\alpha}^L,\;\nabla_{\beta}^L\} = \Gamma_{\alpha\beta}^mA^L_m\;,\;\;
   \{\nabla_{\hat{\alpha}}^R,\;\nabla_{\hat{\beta}}^R\} = 
   \Gamma_{\hat{\alpha}\hat{\beta}}^mA^R_m
\end{equation}
(The existence of such $A^L$ and $A^R$ {\em are} the constraints.) All we need to 
do is to explain how $\nabla_{\alpha}^L$ anticommutes with $\nabla_{\hat{\alpha}}^R$. For that we add a copy of the
finite-dimensional algebra ${\bf g}_{\bar{0}}= so(1,4)\oplus so(5)$ with the generators denoted
$t^0_{[mn]}$. We impose the commutation relations:
\begin{align}
   \{\nabla^L_{\alpha},\;\nabla^R_{\hat{\alpha}}\} \;&= 
   f_{\alpha\hat{\alpha}}{}^{[mn]}t^0_{[mn]}
   \label{QLNablaLNablaR}\\   
   [t^0_{[mn]},\;\nabla^L_{\alpha}] \;&= f_{[mn]\alpha}{}^{\beta}\nabla^L_{\beta}
   \label{QLtNablaL}
   \\   
   [t^0_{[mn]},\;\nabla^R_{\hat{\alpha}}] \;&= f_{[mn]\hat{\alpha}}{}^{\hat{\beta}}
   \nabla^R_{\hat{\beta}}
   \label{QLtNablaR}
   \\  
   [t^0_{[mn]},\;t^0_{[pq]}] \;&= f_{[mn][pq]}{}^{[rs]}t^0_{[rs]}
   \label{QLtt}
\end{align}
where the coefficients $f_{\bullet\bullet}{}^{\bullet}$ are the structure constants of ${\bf g} = {\bf psu}(2,2|4)$.

One can consider the Lie algebra generated by the letters $\nabla^L_{\alpha},\nabla^R_{\hat{\alpha}},t^0_{[mn]}$ with 
the above relations, or the associative algebra generated by them. The 
associative algebra is the same as the universal enveloping $U{\cal L}_{\rm tot}$. 

The algebra $U{\cal L}_{\rm tot}$ is an example of a quadratic-linear algebra. It is a 
{\em filtered} algebra; $F^pU{\cal L}_{\rm tot}$ consists of those elements which can be 
represented by sums of products of $\leq p$ letters. For example $A_m^L\in F^2U{\cal L}_{\rm tot}$. 

One can also define a {\em homogeneous} quadratic algebra $qU{\cal L}_{\rm tot}$ as an 
associative 
algebra generated by the letters $\nabla^L_{\alpha},\nabla^R_{\hat{\alpha}},t^0_{[mn]}$ with the relations (\ref{BasicRelations}) and 
$\{\nabla^L_{\alpha},\nabla^R_{\hat{\beta}}\} = [t^0_{[mn]},\nabla_{\alpha}^L] = [t^0_{[mn]},\nabla_{\hat{\alpha}}^R] = [t^0_{[mn]},t^0_{[pq]}]=0$. The algebra $qU{\cal L}_{\rm tot}$ is
{\em graded}; ${\bf gr}^pU{\cal L}_{\rm tot}$ consists of those elements which consist of $p$ letters. 

\noindent\stepcounter{Theorems}
{\bf Theorem} \arabic{Theorems} {\bf (PBW)}: 
\begin{equation}\label{PBW}
   {\bf gr}^p U{\cal L}_{\rm tot} = {\bf gr}^p(qU{\cal L}_{\rm tot})
\end{equation}
{\bf Proof} uses the fact that $qU{\cal L}_{\rm tot}$ is a Koszul quadratic algebra. We will
give a proof following Section 3.6.8 of \cite{LodayVallette}. We will need some standard 
language, which we will now review. Let $V$ be the vector space generated by 
the letters $\nabla^L_{\alpha},\nabla^R_{\hat{\alpha}},t^0_{[mn]}$. Consider the subspace $R\subset V\otimes V$ generated by the 
following elements (the relations of $qU{\cal L}_{\rm tot}$):
\begin{align}
   & t^0_{[mn]}\otimes t^0_{[pq]} - t^0_{[mn]}\otimes t^0_{[pq]} 
   \\  
   & t^0_{[mn]} \otimes \nabla_{\alpha}^L - \nabla_{\alpha}^L \otimes t^0_{[mn]}
   \\    
   & t^0_{[mn]} \otimes \nabla_{\hat{\alpha}}^R - \nabla_{\hat{\alpha}}^R \otimes t^0_{[mn]}
   \\    
   & \nabla_{\alpha}^L \otimes \nabla_{\hat{\beta}}^R + \nabla_{\hat{\beta}}^R\otimes \nabla_{\alpha}^L
   \\
   & (\Gamma_{m_1\cdots m_5})^{\alpha\beta} \nabla_{\alpha}^L\otimes \nabla_{\beta}^L
   \\   
   & (\Gamma_{m_1\cdots m_5})^{\hat{\alpha}\hat{\beta}} 
   \nabla_{\hat{\alpha}}^R\otimes \nabla_{\hat{\beta}}^R
\end{align}
Notice that $qU{\cal L}_{\rm tot}$ can be defined as the {\em factorspace} of the tensor 
algebra (=free algebra) $TV$ modulo the ideal generated by $R$. 

The dual coalgebra $U{\cal L}_{\rm tot}^{\ashriek}$ is defined as the following {\em subspace} of $TV$:
\begin{equation}
   U{\cal L}_{\rm tot}^{\ashriek} = 
   \;{\bf C}\;\; \oplus \;\;V \;\; \oplus \;\; R \;\; 
   \oplus \;\;
   \bigoplus_{p=3}^{\infty}\bigcap\limits_{q=0}^{p-2} 
   (V^{\otimes q}\otimes R\otimes V^{\otimes (p-q-2)})
\end{equation}
The coalgebra structure is induced from the standard coalgebra structure of
the tensor product:
\begin{equation}
   \Delta(a\otimes b\otimes c \otimes\cdots) =
   a | (b\otimes c \otimes\cdots) + (a\otimes b) | (c\otimes\cdots ) +
   (a\otimes b\otimes c)|(\cdots) + \ldots
\end{equation}

\paragraph     {Explanation of notations:}
For any coalgebra $C$, the coproduct $\Delta$ acts from $C$ to $C\otimes C$. In our case, 
it so happens that $C$ is itself defined as a tensor product. In this case it
is common to use the notation $C|C$ instead of $C\otimes C$, just to avoid
confusion. The spaces $C|C|\cdots |C$ form the so-called {\em cobar complex}, 
because there is a natural differential:
\begin{equation}
   d(x|y|z|\cdots) = \Delta(x)|y|z|\cdots - x|\Delta(y)|z|\cdots +
   x|y|\Delta(z)|\cdots - \ldots
\end{equation}
The nilpotence of this differential is equivalent to the co-associativity of 
$\Delta$. This complex is denoted $\Omega(C)$. As a vector space $\Omega(C)$ is:
\begin{equation}
   \Omega(C) = \bigoplus_{p=0}^{\infty} C^{\otimes p}
\end{equation}
It is naturally an algebra, just a tensor (=free) algebra over $C$:
\begin{equation}
   \Omega(C) = TC
\end{equation}
Also notice that $d$ respects the multiplication: 
$d(X|Y)=d(X)|Y - (-1)^{rk(X)}X|dY$. This means that $\Omega(C)$ is a differential
algebra. Let us consider the cohomology of $d$.

\noindent\stepcounter{Lemmas}
{\bf Lemma \arabic{Lemmas}}:
\begin{equation}
   H_d^0(\Omega(U{\cal L}_{\rm tot}^{\ashriek})) = qU{\cal L}_{\rm tot}
\end{equation}
{\bf Proof:} This is obvious from the definitions. 

So far the definition of $U{\cal L}_{\rm tot}^{\ashriek}$ only used the homogeneous relations of 
$qU{\cal L}_{\rm tot}$, it is really $(qU{\cal L}_{\rm tot})^{\ashriek}$ rather than $U{\cal L}_{\rm tot}^{\ashriek}$. We have to somehow take
into acount the nonzero right hand sides of (\ref{QLNablaLNablaR}), (\ref{QLtNablaL}), (\ref{QLtNablaR}), (\ref{QLtt}). This
is done by supplying $U{\cal L}_{\rm tot}^{\ashriek}$ with a differential $d_1$, which is defined as 
follows\footnote{Notice that the signs do not depend on whether $a,b,c,\ldots$ are ``odd'' 
or ``even'', and in fact we do not use such words at this point. The notion of
``odd'' or ``even'' elements only becomes useful when we say that our 
quadratic-linear algebra is in fact a universal enveloping of a {\em super}-Lie
algebra.}:
\begin{equation}
   d_1 (a\otimes b\otimes c \otimes\cdots) = 
   ((d_1(a\otimes b))\otimes c \otimes\cdots) - 
   (a\otimes(d_1(b\otimes c))\otimes\cdots) + \ldots
\end{equation}
\begin{align}
   d_1 \left( t^0_{[kl]} \otimes t^0_{[mn]} - t^0_{[mn]} \otimes t^0_{[kl]} \right) \;& = \;
   f_{[kl][mn]}{}^{[pq]}t^0_{[pq]}
   \\   
   d_1 \left( t^0_{[mn]} \otimes \nabla_{\alpha}^L - \nabla_{\alpha}^L \otimes t^0_{[mn]} \right) \;& = \;
   f_{[mn]\alpha}{}^{\beta} \nabla_{\beta}^L
   \\   
   d_1 \left( t^0_{[mn]} \otimes \nabla_{\hat{\alpha}}^R - \nabla_{\hat{\alpha}}^R \otimes t^0_{[mn]} \right) \;& = \;
   f_{[mn]\hat{\alpha}}{}^{\hat{\beta}} \nabla_{\hat{\beta}}^R
   \\    
   d_1\left( \nabla_{\alpha}^L \otimes \nabla_{\hat{\beta}}^R + \nabla_{\hat{\beta}}^R\otimes \nabla_{\alpha}^L \right) \;& = \;
   f_{\alpha\hat{\beta}}{}^{[mn]} t^0_{[mn]}
   \\
   d_1\left((\Gamma_{m_1\cdots m_5})^{\alpha\beta} \nabla_{\alpha}^L\otimes \nabla_{\beta}^L\right) \;& = \; 0
   \\
   d_1\left((\Gamma_{m_1\cdots m_5})^{\hat{\alpha}\hat{\beta}} \nabla_{\hat{\alpha}}^R\otimes \nabla_{\hat{\beta}}^R\right) \;& = \; 0
\end{align}
The verification of the nilpotence of $d_1$ is equivalent to the verification of
the Jacobi identity of ${\cal L}_{\rm tot}$ in filtration $\leq 3$. There are the following 
cases to verify:
\begin{align}
   d^2_1\left(
      \Gamma^{\alpha\beta}_{m_1\ldots m_5}
      \nabla^L_{\alpha}\wedge \nabla^L_{\beta}\wedge \nabla^R_{\hat{\alpha}}
   \right)\;=&\; 0
   \label{JacobiLLR}\\   
   d^2_1\left(
      \Gamma^{\alpha\beta}_{m_1\ldots m_5}
      \nabla^L_{\alpha}\wedge \nabla^L_{\beta}\wedge t^0_{[mn]}
   \right)\;=&\; 0
   \label{JacobiLLT}\\ 
   d^2_1\left(
      \nabla^L_{\alpha}\wedge \nabla^R_{\hat{\beta}}\wedge t^0_{[mn]}
   \right)\;=&\; 0
   \label{JacobiLRT}\\   
   d^2_1\left(
      \nabla^L_{\alpha}\wedge t^0_{[mn]}\wedge t^0_{[pq]} 
   \right)\;=&\;0
   \label{JacobiLTT}\\   
   d^2_1\left(
      t^0_{[mn]}\wedge t^0_{[pq]}\wedge t^0_{[rs]} 
   \right)\;=&\;0
   \label{JacobiTTT}
\end{align}
and similar equations with $L\leftrightarrow R$. Eq. (\ref{JacobiTTT}) is the Jacobi identity
for ${\bf g}_{\bar{0}}$. Eq. (\ref{JacobiLTT}) says that the spinor representation is a representation of 
${\bf g}_{\bar{0}}$. Eq. (\ref{JacobiLRT}) is one of the Jacobi identities of the $psl(4|4)$:
\begin{equation}
   f_{\alpha\hat{\beta}}{}^{[pq]}f_{[pq][mn]}{}^{[rs]} =
   f_{\alpha\hat{\gamma}}{}^{[rs]} f_{\hat{\beta}[mn]}{}^{\hat{\gamma}} +
   f_{\alpha[mn]}{}^{\gamma} f_{\gamma\hat{\beta}}{}^{[rs]}
\end{equation}
Eq. (\ref{JacobiLLT}) is derived as follows. After first time applying $d_1$ we get:
\begin{align}
   & d_1\left(
      \Gamma^{\alpha\beta}_{m_1\ldots m_5}
      \nabla^L_{\alpha}\wedge \nabla^L_{\beta}\wedge t^0_{[mn]}
   \right)\;=\;
   \nonumber \\   
   =\;&   \Gamma^{\alpha\beta}_{m_1\ldots m_5} \left(
      f_{\alpha[mn]}{}^{\gamma} \nabla^L_{\gamma}\wedge \nabla^L_{\beta} +
      f_{\beta[mn]}{}^{\gamma}\nabla^L_{\alpha}\wedge \nabla^L_{\gamma}
   \right)\;=\;
   \nonumber \\  
   =\;&   [\Gamma_{mn},\Gamma_{m_1\ldots m_5}]^{\alpha\beta}
   \nabla^L_{\alpha}\wedge \nabla^L_{\beta}
\end{align}
Since $[\Gamma_{mn},\Gamma_{m_1\ldots m_5}]$ is a five-form, the second application of $d_1$ results in
zero. Eq. (\ref{JacobiLLR}) is derived as follows:
\begin{align}
   & d^2_1\left(
      \Gamma^{\alpha\beta}_{m_1\ldots m_5}
      \nabla^L_{\alpha}\wedge \nabla^L_{\beta}\wedge \nabla^R_{\hat{\alpha}}
   \right)\;=
   \nonumber   \\     
   =\;&
   2\Gamma^{\alpha\beta}_{m_1\ldots m_5}
   f_{\beta\hat{\alpha}}{}^{[mn]}f_{\alpha[mn]}{}^{\gamma}\nabla^L_{\gamma} \;=
   \nonumber   \\  
   =\;&
   -\Gamma^{\alpha\beta}_{m_1\ldots m_5}f_{\beta\alpha}{}^kf_{\hat{\alpha}k}{}^{\gamma}
   \nabla^L_{\gamma} \;=\;0
\end{align}
where we have taken into account that $f_{\beta\alpha}{}^k = \Gamma_{\beta\alpha}{}^k$ and therefore the 
contraction with $\Gamma^{\alpha\beta}_{m_1\ldots m_5}$ is zero.

\noindent\stepcounter{Lemmas}
{\bf Lemma \arabic{Lemmas}}: 
\begin{equation}
   H_{d+d_1}^0(\Omega(U{\cal L}_{\rm tot}^{\ashriek})) = U{\cal L}_{\rm tot}
\end{equation}
{\bf Proof:} This is also obvious from the definitions. 

Notice that until now we have not done anything nontrivial, just developed a 
language. But now we are ready to proceed with the proof of the PBW theorem
(\ref{PBW}). Before the proof, we probably have to explain why the statement is 
nontrivial. Let us consider, for example, the following element of ${\cal L}_{\rm tot}$:
\begin{equation}
 X=  [\{\nabla^L_{\alpha},\nabla^L_{\beta}\},\nabla^L_{\gamma}]
\end{equation}
This expression can be represented by the following element 
of $V|V|V\subset\Omega(U{\cal L}_{\rm tot}^{\ashriek})$:
\begin{equation}
   X = (\nabla_{\alpha}^L|\nabla_{\beta}^L + \nabla_{\beta}^L|\nabla_{\alpha}^L)|
   \nabla_{\gamma}^L - 
   \nabla_{\gamma}^L|
   (\nabla_{\alpha}^L|\nabla_{\beta}^L + \nabla_{\beta}^L|\nabla_{\alpha}^L)
\end{equation}
The question is, how do we know that this element is nonzero? Maybe one can
prove that it is zero, using the relations of ${\cal L}_{\rm tot}$? We know however that it is
nonzero as an element of ${\cal L}_L$. (We are not going to prove it now; in fact this 
particular expression corresponds to the field strength superfield.) The ${\cal L}_L$ is
a {\em homogeneous} quadratic algebra. We want to prove that $X$ it is also 
nonzero as an element of ${\cal L}_{\rm tot}$, an {\em inhomogeneous} (quadratic-linear) 
algebra. The danger is that maybe
there is some element $Y_0$, for example in $R|V|V\subset \Omega(U{\cal L}_{\rm tot}^{\ashriek})$, such that 
$dY_0=0$ and $d_1Y_0=X$. This would imply that $(d+d_1)Y_0 = X$ and therefore
$X$ is actually zero as an element of ${\cal L}_{\rm tot}$. Or, perhaps there are 
$Y_0\in V|V|V|R$ and $Y_1\in V|V|R$ such that $d_1 Y_1 = X$ and $dY_1=-d_1Y_0$ and 
$dY_0=0$; then again $(d+d_1)(Y_0+Y_1)=X$ and therefore $X$ is zero. We deal
with such fears in the following manner. Suppose $X=(d+d_1)Y$ and $Y_0$ be the 
highest bar-order term of $Y$ (the term with the highest number of $|$). Then 
$dY_0 = 0$. Because $q{\cal L}_{\rm tot}$ is Koszul\footnote{The Koszul property implies that 
the cohomology group corresponding to $d$-closed $Y_0$ modulo $d$-exact $Y_0$ 
vanishes, see Section 3.6.8 of \cite{LodayVallette} for details. The algebra of 
functions of ten-dimensional spinors satisfying
$(\lambda\Gamma^m\lambda)=0$ is Koszul by the results of 
\cite{Bezrukavnikov:KoszulProperty}. The SYM algebras ${\cal L}_L$  and 
${\cal L}_R$ are both Koszul as quadratic duals to the Koszul algebra of pure 
spinors. The algebra $qU{\cal L}_{\rm tot}$ is the commutative product of 
$U{\cal L}_L$, $U{\cal L}_R$ and $\Lambda{\bf g}_{\bar{0}}$, and therefore is 
Koszul by the Corollary 1.2 in Chapter 3 of \cite{PolishchukPositselski} (where
the commutative product is denoted $\otimes^{q=1}$).}, this implies  that $Y_0=dZ_0$. We therefore
have $(d+d_1)(Y-(d+d_1)Z_0) = X$, and the bar-order of $Y- (d+d_1)Z_0$ is
one less than the bar-order of $Y$. We repeat this until the bar-order of $Y$
is equal to the bar-order of $X$. Now $(d+d_1)Y=X$ implies that the leading
order term in $X$ is zero in $qU{\cal L}_{\rm tot}$. This contradicts the assumption and
completes the proof of the PBW theorem. 

\noindent\stepcounter{Theorems}
{\bf Theorem \arabic{Theorems}}: as a linear space 
\begin{equation}\label{LtotAsLinearSpace}
{\cal L}_{\rm tot}= {\cal L}_L \oplus {\cal L}_R \oplus {\bf g}_{\bar{0}}
\end{equation}
{\bf Proof} The Lie superalgebra ${\cal L}_{\rm tot}$ can be considered a subspace of $U{\cal L}_{\rm tot}$, 
consisting of those elements which can be represented as nested commutators.
Then (\ref{LtotAsLinearSpace}) follows from the PBW theorem.

\paragraph     {Comment:} A physical interpretation of $qU{\cal L}_{\rm tot}$ could be the flat space 
limit of $U{\cal L}_{\rm tot}$.

\subsection{BRST complex}
\subsubsection{The structure of the dual coalgebra}
Besides the PBW theorem, the Koszulity also implies that the complex 
$(U{\cal L}_{\rm tot})^{\ashriek}\otimes U{\cal L}_{\rm tot}$ is acyclic. Notice that $(U{\cal L}_{\rm tot})^{\ashriek}$ has the following structure.
As a linear space:
\begin{align}
   (U{\cal L}_{\rm tot})^{\ashriek} = (U{\cal L}_L)^{\ashriek} \otimes (U{\cal L}_R)^{\ashriek} \otimes \Lambda {\bf g}_{\bar{0}}
\end{align}
Any element $\omega \in (U{\cal L}_{\rm tot})^{\ashriek}$ can be presented as a linear sum of the expressions 
of the ``decomposable'' elements of the form:
\begin{align}
   \omega = \;&   F^{\alpha_1\cdots\alpha_p\;
     \hat{\beta}_1\cdots\hat{\beta}_q\; 
     [m_1n_1]\cdots [m_rn_r]}\;
   \nonumber \\   
   \;& \nabla^L_{\alpha_1}\otimes\cdots\otimes\nabla^L_{\alpha_p}\otimes
   \nabla^R_{\hat{\beta}_1}\otimes\cdots\otimes\nabla^R_{\hat{\beta}_q}\otimes
   t^0_{[m_1n_1]}\otimes\cdots\otimes t^0_{[m_rn_r]}\; +
   \nonumber \\  
   \;& + \; \ldots
   \label{ElementOfULtotAntiShriek}
\end{align}
where $F^{\alpha_1\cdots\alpha_p\;
     \hat{\beta}_1\cdots\hat{\beta}_q\; 
     [m_1n_1]\cdots [m_rn_r]}$ is symmetric in $\alpha_1,\ldots,\alpha_p$ and in $\hat{\beta}_1,\ldots,\hat{\beta}_q$ 
and antisymmetric in $[m_1n_1],\ldots,[m_rn_r]$ and satisfies:
\begin{align}
   \Gamma_{\alpha_1\alpha_2}^k  F^{\alpha_1\cdots\alpha_p\;
     \hat{\beta}_1\cdots\hat{\beta}_q\; 
     [m_1n_1]\cdots [m_rn_r]} \;& = \; 0
   \\   
   \Gamma_{\hat{\beta}_1\hat{\beta}_2}^k  F^{\alpha_1\cdots\alpha_p\;
     \hat{\beta}_1\cdots\hat{\beta}_q\; 
     [m_1n_1]\cdots [m_rn_r]} \;& = \; 0
\end{align}
and $\ldots$ in (\ref{ElementOfULtotAntiShriek}) stand for the terms which are obtained from the first term by
permutations of the tensor product, which are needed so that the resulting
expression belong to the exterior product of $p+q+r$ copies of the linear
superspace generated by $\nabla^L$, $\nabla^R$ and $t^0$. For example, when $p=2$ and $r=1$,
we get:
\begin{align}
   \omega \;& = \; F^{\alpha_1\alpha_2[mn]} \left(
      \nabla^L_{\alpha_1} \otimes \nabla^L_{\alpha_2} \otimes t^0_{[mn]} -
      \nabla^L_{\alpha_1} \otimes t^0_{[mn]} \otimes \nabla^L_{\alpha_2} +
      t^0_{[mn]} \otimes \nabla^L_{\alpha_1} \otimes \nabla^L_{\alpha_2}
   \right)
\end{align}
The action of $d_1$ on this $\omega$ is:
\begin{align}
   d_1\;\omega \;& = \;F^{\alpha_1\alpha_2[mn]}
      f_{[mn]\alpha_1}{}^{\alpha}\nabla^L_{\alpha}\otimes \nabla^L_{\alpha_2} 
\end{align}

\subsubsection{Computing $H^p(U{\cal L}_{\rm tot}\;,\; V)$ as  $\mbox{Ext}_{U{\cal L}_{\rm tot}}({\bf C},V)$}
We have seen that the complex:
\begin{align}
\ldots 
\rightarrow (U{\cal L}_{\rm tot})_2^{\ashriek}\otimes U{\cal L}_{\rm tot}
\rightarrow (U{\cal L}_{\rm tot})_1^{\ashriek}\otimes U{\cal L}_{\rm tot}
\rightarrow U{\cal L}_{\rm tot} \rightarrow {\bf C} \rightarrow 0
\end{align}
provides a free resolution of $\bf C$ as a $U{\cal L}_{\rm tot}$-module. Therefore, for any 
representation $V$ of $U{\cal L}_{\rm tot}$, we can compute the cohomology group:
\begin{align}
   H^p(U{\cal L}_{\rm tot}\;,\; V) = \mbox{Ext}_{U{\cal L}_{\rm tot}}({\bf C},V)
\end{align}
as the cohomology of the complex $\mbox{Hom}_{U{\cal L}_{\rm tot}}\left( 
   (U{\cal L}_{\rm tot})_p^{\ashriek}\otimes U{\cal L}_{\rm tot}\;,\; V
\right)$:
\begin{align}
   0\rightarrow V \rightarrow
   \mbox{Hom}_{\bf C}\left( (U{\cal L}_{\rm tot})^{\ashriek}_1\;,\;V \right)
   \rightarrow
   \mbox{Hom}_{\bf C}\left( (U{\cal L}_{\rm tot})^{\ashriek}_2\;,\;V \right)
   \rightarrow
   \ldots
   \label{MixedPSLieComplex}
\end{align}
We will now interpret this complex in terms of ghosts.
An element of $(U{\cal L}_{\rm tot})^{\ashriek}$ is the sum of expressions of the form (\ref{ElementOfULtotAntiShriek}). Notice that
the dual space $\left((U{\cal L}_{\rm tot})^{\ashriek}\right)'$ is the space of functions of commuting variables 
$c_L^{\alpha}$, $c_R^{\hat{\alpha}}$ satisfying the pure spinor constraints $c_L^{\alpha}\Gamma^m_{\alpha\beta}c_L^{\beta} = c_R^{\hat{\alpha}}\Gamma^m_{\hat{\alpha}\hat{\beta}}c_R^{\hat{\beta}} =0$  and 
anticommuting variables $c_0^{[mn]}$. In this language the BRST operator becomes
\begin{align}
   Q_{\rm BRST} = \;& 
   c_0^{[mn]}\rho(t^0_{[mn]}) + 
   f^{\alpha}{}_{\beta[mn]} c_L^{\beta} c_0^{[mn]} 
   {\partial\over\partial c_L^{\alpha}} \;+ \;
   f^{\hat{\alpha}}{}_{\hat{\beta}[mn]} c_R^{\hat{\beta}} c_0^{[mn]} 
   {\partial\over\partial c_R^{\hat{\alpha}}}\;+
   \nonumber \\  
   &+{1\over 2}\;f^{[mn]}{}_{[m_1n_1][m_2n_2]} c_0^{[m_1n_1]} c_0^{[m_2n_2]}
   {\partial \over \partial c_0^{[mn]}} \;+ 
   \nonumber \\   
   & 
   + c_L^{\alpha}\rho(\nabla^L_{\alpha}) + 
   c_R^{\hat{\alpha}}\rho(\nabla^R_{\hat{\alpha}}) \;+
   \nonumber \\  
   & +\; 
      f^{[mn]}{}_{\alpha\hat{\alpha}}c_L^{\alpha} c_R^{\hat{\alpha}}
   {\partial \over \partial c_0^{[mn]}} 
\end{align}
Notice that the ghosts corresponding to ${\bf g}_{\bar{0}}$ are non-abelian while the ghosts
$c_L^{\alpha}$ and $c_R^{\hat{\alpha}}$ are pure spinors. We will therefore call (\ref{MixedPSLieComplex})  
the ``mixed complex'': it is the pure spinor
BRST complex coupled with the Serre-Hochschild complex of the finite-dimensional
Lie algebra ${\bf g}_{\bar{0}}$.

\subsubsection{Decoupling of the $c_0$-ghosts}\label{sec:DecouplingOfC0}

\paragraph     {In the mixed complex  (\ref{MixedPSLieComplex})}
Let us consider the decreasing filtration by the power of $c_L^{\alpha}$ plus the power of 
$c_R^{\hat{\alpha}}$. The leading term is the cohomology of ${\bf g}_{\bar{0}}$ with values in the functions of 
$(x, c_0, c_L, c_R)$. Let us restrict ourselves to those vertex operators which are
polynomial functions of $x, c_0, c_L, c_R$. The space of such operators splits as an
infinite sum of finite-dimensional representations of ${\bf g}_{\bar{0}}$. Then the cohomology 
sits on the functions which do not depend on $c_0$ and are invariant under the 
action of ${\bf g}_{\bar{0}}$. The resulting complex is the physical BRST complex for the 
unintegrated vertex operators in $AdS_5\times S^5$:

\paragraph     {In the Serre-Hochschild complex of ${\cal L}_{\rm tot}$} 
Similarly, the decoupling of the $c_0$ ghosts in the Serre-Hochschild complex
of ${\cal L}_{\rm tot}$ leads to the relative cohomology:
\begin{equation}
   H^p({\cal L}_{\rm tot}\;,\;(U{\bf g})') \;=\; 
   H^p({\cal L}_{\rm tot}\;,\; {\bf g}_{\bar{0}}\;;\; (U{\bf g})')
\end{equation}

\vspace{20pt}
\noindent This establishes the relation between the BRST cohomology and the 
relative Lie algebra cohomology \cite{Mikhailov:2012uh}.

\subsubsection{Cohomology of the ideal}
Consider the ideal $I\subset {\cal L}_{\rm tot}$ such that:
\begin{equation}
   {\cal L}_{\rm tot}/I = {\bf g}
\end{equation}
By the Shapiro's theorem:
\begin{equation}
   H^p({\cal L}_{\rm tot}\;,\; (U{\bf g})') = H^p(I)
\end{equation}
This helps to identify various supergravity field strengths \cite{Mikhailov:2012uh}. 

\subsection{Integrated vertex}
\subsubsection{Generalized Lax operator}
Consider a classical string solution in $AdS_5\times S^5$, {\it i.e.} a field 
configuration in the worldsheet sigma-model solving the classical equations of 
motion. It was shown in \cite{Mikhailov:2013vja,Chandia:2013kja} that one can construct the {\em Lax pair}:

\begin{align}
   L_+  = \;& 
\left(
   {\partial\over \partial \tau^+} 
   + J_{0+}^{[mn]} t^0_{[mn]} 
\right) 
+ J_{3+}^{\alpha} \nabla^L_{\alpha} + J_{2+}^m A^L_m + (J_{1+})_{\alpha} W_L^{\alpha} +
 \nonumber \\   
& 
+ \lambda_L^{\alpha} w^L_{\beta +} \left(
   \{  \nabla^L_{\alpha} \;,\; W_L^{\beta}\}
   - f_{\alpha}{}^{\beta}{}^{[mn]} t^0_{[mn]}
\right)
\label{LPlus} \\     
L_- = \;&
\left(
   {\partial\over\partial\tau^-}
   + J_{0-}^{[mn]} t^0_{[mn]}
\right) 
+ J_{1-}^{\dot{\alpha}} \nabla^R_{\dot{\alpha}} 
+ J_{2-}^m A^R_m + (J_{3-})_{\dot{\alpha}} W_R^{\dot{\alpha}} +
\nonumber \\  
& 
+ \lambda_R^{\dot{\alpha}} w^R_{\dot{\beta}-}\left(
   \{ \nabla^R_{\dot{\alpha}} \;,\; W_R^{\dot{\beta}} \}
   - f_{\dot{\alpha}}{}^{\dot{\beta}}{}^{[mn]} t^0_{[mn]}
\right)
\label{LMinus}
\end{align}
where $J_{\pm}$ and $\lambda, w$ are worldsheet fields and $t_0, \nabla, A, W$ generators of ${\cal L}_{\rm tot}$,
satisfying the zero curvature equations:
\begin{equation}
   [L_+,L_-] = 0
\end{equation}
and having simple BRST transformation laws:
\begin{equation}\label{BRSTCovariance}
   Q_{BRST}L_{\pm} \;=\; 
\left[
   L_{\pm}\;,\;\left(
      \lambda_L^{\alpha}\nabla^L_{\alpha} + 
      \lambda_R^{\dot{\alpha}} \nabla^R_{\dot{\alpha}}
   \right)
\right]
\end{equation}
We will denote $J_{\pm}$ the connections in $L_{\pm}$:
\begin{equation}
   L_{\pm} = {\partial\over\partial \tau^{\pm}} + J_{\pm}
\end{equation}

\subsubsection{Bicomplex $d + Q_{\rm BRST}$}
Let us denote: $J = J_+ d\tau^+ + J_- d\tau^-$ --- an ${\cal L}_{\rm tot}$-valued one-form on the 
worldsheet. For the purpose of calculations, it is convenient to assume that
$d\tau^+$ and $d\tau^-$ anticommute with the worldsheet fields $\theta$:
\begin{align}
   d\tau^+ \theta^{\alpha} = \;& - \theta^{\alpha} d\tau^+
   \\    
   d\tau^- \theta^{\alpha} = \;& - \theta^{\alpha} d\tau^-
\end{align}
We also introduce arbitrarily many anticommuting parameters $\epsilon_1, \epsilon_2,\ldots$, which 
anticommute among themselves, with $\theta$, and with $d\tau^{\pm}$. With these 
notations, we have:
\begin{align}
&   
\left(\epsilon_1 d + \epsilon_1 Q_{\rm BRST}\right)
\left(\epsilon_2 d\tau^j J_j - \epsilon_2 \lambda\right)\;=
\\   
=\;& -{1\over 2} 
\left[
   \;\epsilon_1 d\tau^i J_i - \epsilon_1\lambda\;\;,\;
   \;\epsilon_2 d\tau^i J_i - \epsilon_2\lambda\;
\right]
\end{align}
Schematically: 
\begin{equation}\label{BicomplexSchematically}
\epsilon_1(d+Q_{\rm BRST})\;\epsilon_2(J-\lambda) = 
- {1\over 2}\left[\;\epsilon_1(J-\lambda)\;,\;\epsilon_2(J-\lambda)\;\right]
\end{equation}
Also:
\begin{align}\label{BicomplexOnG}
   \epsilon_1 (d+Q_{\rm BRST}) g = - \epsilon_1\pi(J-\lambda)\;g
\end{align}
Given an $n$-cochain $\psi\in C^n({\cal L}_{\rm tot},{\bf g}_{\bar{0}};(U{\bf g})')$, let us consider an inhomogeneous 
form $\mbox{ev}(\psi)$ on the worldsheet which can schematically be defined by the 
following formula: 
\begin{align}
   \mbox{ev}_{\epsilon_1,\ldots,\epsilon_n}(\psi) = 
   \psi\left(
      \epsilon_1(J - \lambda)\otimes \epsilon_2(J - \lambda)\otimes\ldots
   \right) (g)
\end{align}
This schematic notation is deciphered as follows. Notice that $\psi$ is a function 
of the type:
\begin{align}
   \Lambda^n {\cal L}_{\rm tot} \rightarrow [ U{\bf g} \rightarrow {\bf C}]
\end{align}
We first evaluate it on $\epsilon_1(J-\lambda)\otimes \epsilon_2(J-\lambda)\otimes\ldots$, which gives us a function 
of the type $U{\bf g} \rightarrow {\bf C}$. We then evaluate it on a ``group element''$g\in U{\bf g}$. 
The ``group elements'' are defined as expressions of the form $g = e^{\xi}$ where 
$\xi\in {\bf g}$. Being infinite series, they strictly speaking do not belong to $U{\bf g}$. 
This rises the question of convergence, which we will ignore.

Then we observe:
\begin{align}
   \epsilon_1(d + Q_{\rm BRST})\;\mbox{ev}_{\epsilon_2,\ldots,\epsilon_{n+1}}(\psi) 
   = \; {1\over n+1}\;\mbox{ev}_{\epsilon_1,\ldots,\epsilon_{n+1}}(Q_{\rm Lie}\psi)
\end{align}
The derivation of this formula, besides (\ref{BicomplexSchematically}) and (\ref{BicomplexOnG}), also uses the fact 
that $\psi$ is a {\em relative} cocycle, and therefore:
\begin{align}
\psi\left(\{\;\lambda\;,\lambda\;\}\;\otimes\;\ldots\right)\; = \;
\psi\left(
   2\lambda_L^{\alpha}\lambda_R^{\hat{\alpha}}f_{\alpha\hat{\alpha}}{}^{[mn]}t^0_{[mn]}
   \;\otimes\;\ldots \;
\right)\; = \;0
\end{align}
In our case $\psi$ is a 2-cocycle. Therefore:
\begin{equation}
   (d+Q_{\rm BRST}) \mbox{ev}_{\epsilon_1,\epsilon_2}(\psi) = 0
\end{equation}
This means that the ghost number two part of $\mbox{ev}_{\epsilon_1,\epsilon_2}(\psi)$ is an unintegrated
vertex, and the ghost number zero part  of $\mbox{ev}_{\epsilon_1,\epsilon_2}(\psi)$ is an integrated vertex.

Therefore our construction provides one way of thinking about the relation
between unintegrated and integrated vertices.
\ifodd\amshow
\hspace{-10pt}
\includegraphics[scale=0.5]{snapshots/definition-of-Lie-Algebra-Cohomology.png}\\
\hspace{-10pt}
\includegraphics[scale=0.25]{photos/comparison-with-Fuks.png}\\
\fi

\section{General curved superspace}\label{sec:GeneralCurved}

The pure spinor description of the Type IIB SUGRA emphasizes the local Lorentz
symmetry of the supergravity theory. More specifically, the Type IIB 
superstring combines left and right sectors, and there are two copies of the
local Lorentz group. 

We will now describe some structure on the superspace, which we call ``SUGRA 
data''. We first describe it as an abstract geometrical structure, and then
explain how it emerges in the sigma-model.

\subsection{Weyl superspace}\label{sec:WeylSuperspace}
The formulation of  the pure spinor sigma-model in \cite{Berkovits:2001ue} uses the so-called Weyl 
superspace. In this formalism, besides the local Lorentz symmetry, there are 
also two copies of the local dilatation symmetries:
\begin{equation}
   \hat{{\bf h}} = \hat{{\bf h}}_L \oplus \hat{{\bf h}}_R = spin(1,9)_L\oplus {\bf R}_L\oplus spin(1,9)_R\oplus {\bf R}_R
\end{equation}
Let $S_L\oplus S_R$ denote the spinor representation of $\hat{\bf h}$. We will also denote $\hat{H}$ the
Lie group corresponding to $\hat{\bf h}$. To summarize:
\begin{align}
   \hat{H} \;=&\;
   Spin(1,9)_L\times {\bf R}^{\times}_L\times Spin(1,9)_R\times {\bf R}^{\times}_R
   \\    
   \hat{\bf h} \;=&\; \mbox{Lie}(H)
   \\   
   S_L\oplus S_R \;=&\; \mbox{\tt\small spinor representation of}\; H
\end{align}
We will now start describing the SUGRA data. 

\vspace{12pt}
\noindent
Let $M$ be a $10|32$-dimensional supermanifold, the super-space-time.

\paragraph     {The first part of the SUGRA data} is:
\begin{itemize}
\item a distribution ${\cal S}_L \oplus {\cal S}_R \subset TM$
\item for every point $x\in M$, an orbit of the action of $\hat{H}$ on some linear map
   ${\cal D}\;:\;S_L\oplus S_R\to {\cal S}_L(x)\oplus {\cal S}_R(x)$ (the action of $h\in\hat{H}$ is ${\cal D}\mapsto {\cal D}\circ h$); 
   notice that the map $\cal D$ itself does not enter into the SUGRA data, only its 
   orbit (with the action of $\hat{H}$ on it)
\end{itemize}
Let $\widehat{M} \xrightarrow{\pi} M$ be the principal bundle over $M$ whose fiber over a point $x\in M$
is that orbit. In other words, a point of $\widehat{M}$ is a pair $(x,{\cal D})$. Let $\pi$ denote 
the natural projection:
\begin{align}
   &\pi\;:\; \widehat{M}\to M
   \nonumber \\ 
   &\pi(x,{\cal D}) \;= x
   \label{Projection}
\end{align}
More explicitly, any linear map $\cal D$ is of the form:
\begin{align}
   {\cal D}(s_L+s_R) \;&= s_L^{\alpha}E^L_{\alpha} + s_R^{\hat{\alpha}}E^R_{\hat{\alpha}}
   \label{DLandDR}\\   
   E^L_{\alpha}\;&\in {\cal S}_L(x)
   \\   
   E^R_{\hat{\alpha}}\;&\in {\cal S}_R(x)
\end{align}
Sometimes we will simply write $E_{\alpha}$ and $E_{\hat{\alpha}}$ instead of $E^L_{\alpha}$ and $E^R_{\hat{\alpha}}$.

Let $\mbox{Vect}(\widehat{M})=\Gamma(T\widehat{M})$ denote the infinite-dimensional space of all vector 
fields on $\widehat{M}$. 

\paragraph     {The second part of the SUGRA data} is a map
\begin{align}
   & {\bf D}\;:\; S_L\oplus S_R \to \mbox{Vect}(\widehat{M})
   \\  
   & {\bf D}(s_L + s_R) \;=\; s_L^{\alpha}{\bf D}^L_{\alpha} + s_R^{\hat{\alpha}}{\bf D}^R_{\hat{\alpha}}
   \label{DefDComponents}
\end{align}
satisfying the following properties:
\begin{itemize}
\item $\bf D$ commutes with the action of $\hat{H}$
\item $\bf D$ is ``fixed modulo $\mbox{Vect}\widehat{M}/M$'' in the following sense: for any
   point $(x,{\cal D})\in M$ let $\pi(x)$ be the natural projection  $T_{(x,{\cal D})}\widehat{M}\to T_xM$, then 
\begin{equation}
   \pi(x)\Big(({\bf D}(s_L+s_R))(x,{\cal D})\Big) = {\cal D}(s_L+s_R)
\end{equation}
(in other words, only the vertical component of $\bf D$ is non-obvious; the 
projection to $TM$ is tautological)
\item SUGRA constraints:
\begin{align}
   &\{{\bf D}(s_L+s_R)\;,\;{\bf D}(s_L+s_R)\} = 
   \nonumber \\  
   =\;&(s_L\Gamma^m s_L){\bf A}^L_m + (s_R\Gamma^m s_R){\bf A}^R_m \;+ 
   \nonumber \\  
   & \;+
   R^{LL}_{\alpha\beta}s_L^{\alpha}s_L^{\beta} +
   R^{RR}_{\dot{\alpha}\dot{\beta}}s_R^{\dot{\alpha}}s_R^{\dot{\beta}} +
   R^{LR}_{\alpha\dot{\beta}} s_L^{\alpha} s_R^{\dot{\beta}}
   \label{SpecialD}
\end{align}
where:
\begin{itemize}
\item ${\bf A}_m^L$ and ${\bf A}_m^R$ are some sections of ${\cal T}\widehat{M}$ and
\item $R_{\alpha\beta}^{LL}$, $R_{\dot{\alpha}\dot{\beta}}^{RR}$ and $R_{\alpha\dot{\beta}}^{LR}$ some sections of ${\cal T}\widehat{M}/M$ 
   ({\it i.e.} vertical vector fields); they are essentially ``curvatures''
\end{itemize}
Notice that satisfying the SUGRA constraints {\em does} depend on the vertical 
component of ${\bf D}$.
\end{itemize}
Moreover:
\begin{itemize}
   \item there is an equivalence relation, which we will describe in Section \ref{sec:ShiftGaugeTransformations}
\end{itemize}

\subsection{Relation to the formalism of \cite{Berkovits:2001ue}}
\subsubsection{SUGRA constraints, oversimplified}\label{sec:OverSimplified}
Let $M$ be the super-space-time. In supergravity, $M$ comes equipped with the
distribution ${\cal S}\subset TM$. The SUGRA constraints are conditions on the Frobenius
form of ${\cal S}$, which go roughly speaking as follows. We choose some vector fields
$\nabla_{\alpha}$, $\alpha\in \{1,\ldots,\mbox{dim}{\cal S}\}$ tangent to ${\cal S}$ and say that:
\begin{equation}\label{SimplifiedConstraints}
   \{\nabla_{\alpha},\nabla_{\beta}\} = \Gamma^m_{\alpha\beta}A_m \mbox{ mod } {\cal S}
\end{equation}
where $A_m$ are some other vector fields. (The point of the constraint being 
that the RHS is proportional to $\Gamma_{\alpha\beta}^m$.) It is important to remember that when
we write such conditions, we need to fix a basis of ${\cal S}$, {\it i.e.} a set
of $\nabla_{\alpha}$. If we choose some linear combination:
\begin{equation}
   \nabla'_{\alpha} = X_{\alpha}^{\beta}\nabla_{\beta}
\end{equation}
then, generally speaking, $\nabla'_{\alpha}$ will not satisfy the constraint (\ref{SimplifiedConstraints}). If we
want  $\nabla'_{\alpha}$ to satisfy the constraint, we should require that $X\in so(1,9)\oplus {\bf R}$ 
--- an antisymmetric matrix plus a scalar. This means that $\cal S$ actually comes 
with an additional structure, namely an orbit of the action of $SO(1,9)\times {\bf R}_{\times}$ 
on some linear map ${\cal D}\;:\; S\to {\cal S}$ where $S$ is the spinor representation of 
$so(1,9)\oplus {\bf R}$. As we said in Section \ref{sec:WeylSuperspace}, the map $\cal D$ itself does not enter into 
the SUGRA data, only its orbit (with the action of $SO(1,9)\times {\bf R}_{\times}$ on it). 
Given a point $x\in M$, and an orbit of $SO(1,9)\times {\bf R}_{\times}$ in ${\cal S}(x)$, we can choose
a point ${\cal D}$ in this orbit, then choose {\em any} set of vector fields $\nabla_{\alpha}$ such 
that $\nabla_{\alpha}(x) = {\cal D}_{\alpha}$, and verify Eq. (\ref{SimplifiedConstraints}). 

This means that it is useful instead of $M$ to consider $\widehat{M}$, which is
the $SO(1,9)\times {\bf R}_{\times}$-bundle over $M$ whose fiber over $x\in M$ is that  
$SO(1,9)\times {\bf R}_{\times}$-orbit in $\mbox{Hom}_{\bf C}(S,{\cal S}(x))$ which we should have received as part 
of our SUGRA data. It is natural to think that the matter fields live in $\widehat{M}$ 
rather than $M$, except that {\em the fiber is a gauge degree of freedom}. 
The fiber can be gauged away because, as we said, the map ${\cal D}\in\mbox{Hom}_{\bf C}(S,{\cal S}(x))$ 
itself does not enter into the SUGRA data, only its orbit. This is how the 
$AdS_5\times S^5$ sigma-model is formulated \cite{Berkovits:2000fe}. In that case $M$ is
$PSU(2,2|4)/(SO(1,4)\times SO(5))$ and $\widehat{M}$ is $PSU(2,2|4)$. The sigma model has 
the $SO(1,4)\times SO(5)$ gauge symmetry which gauges away the fiber. It is 
$SO(1,4)\times SO(5)$  rather than $SO(1,9)\times {\bf R}_{\times}$ because in that particular 
case some of the gauge symmetry can be canonically fixed.

In the sigma-model we couple matter fields with the ghosts $\lambda$ which belong to 
the {\em pure spinor cone} $C\subset S$. As  ${\cal D}\in\mbox{Hom}_{\bf C}(S,{\cal S})$ can be thought of
as linear functions from $S$ to $\cal S$, it make sense to apply it to $\lambda\in S$. The
resulting vector field ${\cal D}(\lambda)$ describes the action of the BRST operator on the
matter fields:
\begin{equation}\label{SimplifiedQvsD}
   Q_{\rm matter} = {\cal D}(\lambda)
\end{equation}

\subsubsection{Sigma-model}
The target space of the sigma-model is $\widehat{M}$, but as we explained there is a 
gauge symmetry which reduces $\widehat{M}\to M$. The action, copied from \cite{Berkovits:2000fe}, is:
\begin{multline}
S = \frac{1}{2\pi\alpha'} \int d^2 z \bigg( {1\over 2} \big(G_{MN}(Z) 
+ B_{MN}(Z)\big) \partial Z^M \overline\partial Z^N 
+ E^\alpha_M (Z) d_\alpha \overline\partial Z^M + \bigg.\\
\qquad + E^{\hat\alpha}_M (Z) \tilde{d}_{\hat\alpha} \partial Z^M 
+ \Omega_{M\alpha}{}^\beta(Z) \lambda^\alpha  w^L_\beta \overline\partial Z^M 
+ \hat\Omega_{M\hat\alpha}{}^{\hat\beta}(Z) \tilde\lambda^{\hat\alpha} 
w^R_{\hat\beta} \partial Z^M \; + \\
\qquad + P^{\alpha\hat\beta}(Z) d_\alpha \tilde d_{\hat\beta} 
+ C^{\beta\hat\gamma}_{\alpha}(Z) \lambda^{\alpha}  w^L_{\beta} \tilde d_{\hat\gamma} 
+ \hat C^{\hat\beta\gamma}_{\hat\alpha}(Z) \tilde\lambda^{\hat\alpha} 
w^R_{\hat\beta} d_{\gamma} \; + \\
+ S^{\beta\hat\delta}_{\alpha\hat\gamma}(Z) 
\lambda^{\alpha}  w^L_{\beta} \tilde \lambda^{\hat\gamma} w^R_{\hat\delta} 
+ {1\over 2} \alpha' \Phi(Z)r \; + \\
\bigg.\qquad  + w^L_{\alpha+}\partial_-\lambda_L^{\alpha} + 
w^R_{\hat{\alpha}-}\partial_+\lambda_R^{\hat{\alpha}}
\bigg)
\label{ActionOfBerkovitsAndHowe}
\end{multline}
In a generic background, one can integrate out $d,\tilde{d}$ and get a 
simpler-looking action. It is postulated that the field $d$ should be the 
same as the density of the BRST charge. This is, essentially, a restriction
on the choice of fields. Notice that the form of the Lagrangian (\ref{ActionOfBerkovitsAndHowe}) is not
invariant under the field redefinitions, specifically under those redefinitions
which mix the ghosts $\lambda$ with the matter fields $Z^M$. (And this, in our opinion, 
is a defect of the formalism in its current form.)

The phase space of this sigma-model will be denoted ${\cal X}$. It can be identified 
with the moduli space of all classical solutions:
$$
   {\cal X} = {
     \mbox{\tt\small the space of classical solutions of the string $\sigma$-model}
   }
$$
We can also consider the space of off-shell field configurations:
$$
{\cal X}_{\rm OS} = 
\mbox{\tt\small the space of off-shell field configurations}
$$

\subsubsection{From pure spinor $Q$ to SUGRA constraints}\label{sec:FromQtoSUGRA}
We just said that the target space of the sigma-model is $\widehat{M}$. This, however, is
not the full truth, because there are also ghosts. With ghosts, the target 
space is a cone in the associated vector bundle of the principal bundle $\widehat{M}$ 
corresponding to the spinor representation of $\hat{H}$:
\begin{equation}\label{TargetSpaceWithGhosts}
   \mbox{\tt\small Target space with ghosts} = 
   \widehat{M}\times_{\hat{H}} (C_L\times C_R)
\end{equation}
where $C_L$ is the pure spinor cone in $S_L$ and $C_R$ the pure spinor cone in $S_R$.
The BRST operator of the sigma-model is a nilpotent odd vector field:
\begin{equation}
Q\in \mbox{Vect}(\widehat{M}\times_{\hat{H}} (C_L\times C_R))
\end{equation}
Generally speaking, consider a coset space $X/G$, where the action of $G$ on $X$
is free and transitive. Then vector fields on $X/G$ can be described as 
follows. Let us start by considering the subalgebra $A\subset\mbox{Vect}(X)$ which 
consists of those vector fields which are invariant under the action of $G$, 
{\it i.e.} $g_*v = v$ for any $g\in G$ (a.k.a ``Atiyah algebroid''). One can check 
that the vertical vector fields (those which are tangent to the orbits of $G$) 
are an ideal $I\subset A$. The factoralgebra is isomorphic to the algebra of vector 
fields on $X/G$:
\begin{equation}\label{VectorFieldsOnFactorspace}
   \mbox{Vect}(X/G)\simeq A/I
\end{equation}
Let us see how this description works in the particular case:
$$X = \widehat{M}\times (C_L\times C_R)\;,\;\; G = \hat{H}\; \mbox{ \tt\small and } X/G = \widehat{M}\times_{\hat{H}} (C_L\times C_R)$$ 
Let us fix {\em some lift}
\begin{equation}
   {\bf Lift}: \widehat{M}\times_{\hat{H}} (C_L\times C_R) \;\rightarrow \;
   \widehat{M}\times (C_L\times C_R)
\end{equation}
Consider ${\bf Lift}_* Q$ --- a vector field on the image of $\bf Lift$. Notice that 
$({\bf Lift}_* Q)^2=0$, but don't forget that  ${\bf Lift}_* Q$ is not a vector field on the 
whole $\widehat{M}\times (C_L\times C_R)$, but only on a submanifold --- the image of $\bf Lift$. 
However, we can define an $\hat{H}$-invariant vector field on the whole  
$\widehat{M}\times (C_L\times C_R)$ using the fact that $\widehat{M}\times (C_L\times C_R)$ is foliated by the
translations of the image of $\bf Lift$ by elements of $\hat{H}$:
\begin{equation}
   \widehat{M}\times (C_L\times C_R) = \bigcup_{h\in \hat{H}} h(\mbox{im}({\bf Lift}))
\end{equation}
This means that we can extend  ${\bf Lift}_* Q$ to the whole $\widehat{M}\times (C_L\times C_R)$ in an
$h$-invariant matter, simply by translating. In other words, let $Q^{\uparrow}$ be the 
nilpotent vector field on $\widehat{M}\times (C_L\times C_R)$ such that:
\begin{align}
   Q^{\uparrow}|_{\mbox{\small im}({\bf Lift})} =\;& {\bf Lift}_* Q  
   \\    
   \;\mbox{ \small\tt for any } h\in\hat{H}\;:\;\;h_*Q^{\uparrow} =\;& Q^{\uparrow} 
\end{align}
A different choice of $\bf Lift$ will result in another $Q^{\uparrow}$, but the difference in 
$Q^{\uparrow}$ will be in adding a vertical vector field, {\it i.e.} and element of $I$. 
This is precisely (\ref{VectorFieldsOnFactorspace}).

Notice that the vertical component of $Q^{\uparrow}$ is $\hat{H}$-invariant. Moreover, 
$C_L\times C_R$ is an orbit of the action of $\hat{H}$ on $S_L\times S_R$. In other words, any 
pair of pure spinors $(\lambda_L,\lambda_R)$ can be obtained from a fixed pair $(\lambda_L^{(0)},\lambda_R^{(0)})$ 
by the action of some element $h\in \hat{H}$. Therefore exists a vertical vector 
field $\omega$ such that the following vector field on $\widehat{M}\times (C_L\times C_R)$:
\begin{equation}
   \widehat{Q} = Q^{\uparrow} + \omega
\end{equation}
acts trivially on $C_L\times C_R$. In other words, $\widehat{Q}$ is a vector field on $\widehat{M}$. 

To clarify the construction, let us describe it in coordinates. A point of
$\widehat{M}\times (C_L\times C_R)$ is described in coordinates as follows: 
$(Z,(E^{\rm L}_{\alpha}), (E^{\rm R}_{\hat{\alpha}}), \lambda_L, \lambda_R)$. A point of $\widehat{M}\times_{\hat{H}} (C_L\times C_R)$ is described in the same 
way but with the equivalence relation:
\begin{equation}\label{EquivalenceRelation}
 (Z,(E^{\rm L}_{\alpha}), (E^{\rm R}_{\hat{\alpha}}), \lambda_L, \lambda_R)
\sim 
 (Z,
 ((h^{-1}_{\rm L})^{\alpha'}_{\alpha}E^{\rm L}_{\alpha'}), 
 ((h^{-1}_{\rm R})^{\hat{\alpha}'}_{\hat{\alpha}}E^{\rm R}_{\hat{\alpha}'}), 
 h_{\rm L}\lambda_L, 
 h_{\rm R}\lambda_R) 
\end{equation}
Our $\bf Lift$ is essentially gauge fixing. It is described by specifying the 
functions:
\begin{equation}\label{GaugeFixing}
   E^{\rm L}_{\alpha} = E^{\rm L0}_{\alpha}(Z)\;\mbox{ \tt\small and }\;
   E^{\rm R}_{\hat{\alpha}} = E^{\rm R0}_{\hat{\alpha}}(Z)
\end{equation}
The way it works, for every point in $\widehat{M}\times_{\hat{H}} (C_L\times C_R)$, to calculate its lift
we use the equivalence relations (\ref{EquivalenceRelation}) to bring its coordinates 
$(Z,(E^{\rm L}_{\alpha}), (E^{\rm R}_{\hat{\alpha}}), \lambda_L, \lambda_R)$ to the form satisfying (\ref{GaugeFixing}). The resulting 
$(Z,(E^{\rm L0}_{\alpha}(Z)), (E^{\rm R0}_{\hat{\alpha}}(Z)), \lambda^{\rm new}_L, \lambda^{\rm new}_R)$ specifies a point in $\widehat{M}\times (C_L\times C_R)$, 
which is the lift. The lift of the BRST field ${\bf Lift}_* Q$ is of the form:
\begin{equation}
     {\bf Lift}_* Q_L = \lambda^{\alpha}_L \left(
        E^{{\rm L0}M}_{\alpha}(Z) {\partial\over\partial Z^M} + 
        X_{\alpha}{}^{\beta}_{\gamma}(Z) 
        \lambda_L^{\gamma}{\partial\over\partial \lambda_L^{\beta}}
        \right)
\end{equation}
We must stress that ${\bf Lift}_* Q$ is only defined on the image of $\bf Lift$. To extend
this vector field to the whole $\widehat{M}\times (C_L\times C_R)$, we must relax the gauge fixing
(\ref{GaugeFixing}). We observe that any $E^{\rm L}_{\alpha}$ and $E^{\rm R}_{\hat{\alpha}}$ can be presented in the form:
\begin{equation}
   E^{\rm L}_{\alpha} = (g^{\rm L})_{\alpha}^{\alpha'}E^{\rm L0}_{\alpha'}(Z)
   \;\mbox{ \tt\small and }\;
   E^{\rm R}_{\hat{\alpha}} = (g^{\rm R})_{\hat{\alpha}}^{\hat{\alpha}'}
   E^{\rm R0}_{\hat{\alpha}'}(Z)
\end{equation}
Let us use $(Z,g^{\rm L}, g^{\rm R}, \lambda_L, \lambda_R)$ as coordinates on $\widehat{M}\times (C_L\times C_R)$. Then we have:
\begin{equation}
   Q_L^{\uparrow} = \lambda^{\alpha}_L \left(
      (g^{\rm L})^{\alpha'}_{\alpha}
      E^{{\rm L0}M}_{\alpha'}(Z) {\partial\over\partial Z^M} + 
      (g^{\rm L})^{\alpha'}_{\alpha}
      X_{\alpha'}{}^{\beta'}_{\gamma'}(Z) 
      ((g^{\rm L})^{-1})^{\beta}_{\beta'}
      (g^{\rm L})^{\gamma'}_{\gamma}
      \lambda_L^{\gamma}{\partial\over\partial \lambda_L^{\beta}}
   \right)
\end{equation}
Finally:
\begin{equation}
   \widehat{Q}_L = \lambda^{\alpha}_L \left(
      (g^{\rm L})^{\alpha'}_{\alpha}
      E^{{\rm L0}M}_{\alpha'}(Z) {\partial\over\partial Z^M} + 
      (g^{\rm L})^{\alpha'}_{\alpha}
      X_{\alpha'}{}^{\beta'}_{\gamma'}(Z)
      (g^{\rm L})^{\gamma'}_{\delta} 
      {\partial\over\partial (g^{\rm L})^{\beta'}_{\delta} }
\right)
\end{equation}
--- a vector field on $\widehat{M}$.

Notice that $\widehat{Q}_L$ depends linearly on $\lambda_L$. Therefore, $\widehat{Q}$ defines sixteen vector 
fields ${\bf D}^L_{\alpha}$:
\begin{equation}
   \widehat{Q}_L = \lambda_L^{\alpha}{\bf D}^L_{\alpha}
\end{equation}
These are the vector fields which were postulated in Section \ref{sec:WeylSuperspace}. 

\paragraph     {Ambiguity}
However, the definition of $\widehat{Q}_L$, and therefore of ${\bf D}^L_{\alpha}$, contains an ambiguity.
It is possible to add to $\widehat{Q}_L$ a vertical vector field:
\begin{equation}
   \widehat{Q}_{L,\rm new} = \widehat{Q}_L + \lambda_L^{\alpha}\omega^{\rm L}_{\alpha}
\end{equation}
such that $\lambda_L^{\alpha}\omega^{\rm L}_{\alpha} \in \mbox{St}(\lambda_L)\subset \hat{\bf h}_L$. This corresponds to the ``shift gauge 
transformations'' of \cite{Berkovits:2001ue}. We will now describe such $\omega_{\alpha}$.

\subsubsection{Shift gauge transformations}\label{sec:ShiftGaugeTransformations}
Let us modify ${\bf D}_{\alpha}^L$ by adding to it a vector field in $T\widehat{M}/M$ ({\it i.e.} 
tangent to the fiber) of the form ({\it cf.}  Eq. (61) of  \cite{Berkovits:2001ue}):
\begin{equation}
   (\omega_{\alpha}^L)^{\beta}_{\gamma} = (\Gamma^n\Gamma^m)^{\beta}_{\gamma} \Gamma^m_{\alpha\bullet}h_L^{\bullet n }
\end{equation}
The characteristic property of such $\omega^{\rm L}_{\alpha}$ is that  $\lambda_L^{\alpha}\omega^{\rm L}_{\alpha} \in \mbox{St}(\lambda_L)\subset \hat{\bf h}_L$; in 
other words:
\begin{equation}
   \lambda_L^{\alpha}(\omega^{\rm L}_{\alpha})^{\beta}_{\gamma}\lambda_L^{\gamma} = 0
\end{equation}
Similarly, we can modify ${\bf D}_{\hat{\alpha}}^R$ by adding to it some $\omega_{\hat{\alpha}}^R$ defined in a similar way;
we stress that $\omega^L$ takes values in $\hat{\bf h}_L$ and $\omega_R$ takes values in $\hat{\bf h}_R$.
Obviously, these ``shift transformations'' depend on two parameters: $h_L^{\alpha n}$ and 
$h_R^{\hat{\alpha} n}$. In terms of Section \ref{sec:WeylSuperspace} this modifies ${\bf D}$:
\begin{equation}
   {\bf D}_{\rm new}(s_L + s_R) = {\bf D}(s_L + s_R) + s_L^{\alpha} \omega_{\alpha}^L
   + s_R^{\hat{\alpha}}\omega_{\hat{\alpha}}^R
\end{equation}

\subsubsection{SUGRA fields}
The action (\ref{ActionOfBerkovitsAndHowe}) involves various SUGRA fields, which are either sections of 
associated vector bundles over $M$, or connections on them. 
They enter the action through their pullback on the string worldsheet.

\paragraph     {Sections}
For example, $P^{\alpha\hat{\alpha}}$ is a section of $\widehat{M}\times_{\hat{H}} (S_L\otimes S_R)\stackrel{\pi}{\rightarrow}M$. Such sections can 
be interpreted as $\hat{H}$-invariant maps\footnote{Indeed, every such map defines
$\sigma:\;M\to\widehat{M}\times_{\hat{H}} (S_L\otimes S_R)$ such
that $\pi\circ\sigma=\mbox{id}$} $\widehat{M}\to S_L\otimes S_R$. From this point of view
we consider $P^{\alpha\hat{\alpha}}$ as a function $P^{\alpha\hat{\alpha}}(Z,(E^{\rm L}_{\beta}), (E^{\rm R}_{\hat{\beta}}))$ such that:
\begin{equation}
   P^{\alpha\hat{\alpha}}(Z,\;((h_{\rm L})_{\beta}^{\beta'}E^{\rm L}_{\beta'}), \;
   ((h_{\rm L})_{\hat{\beta}}^{\hat{\beta}'}E^{\rm R}_{\hat{\beta}'}))\;=\;
   (h_{\rm L}^{-1})^{\alpha}_{\alpha'}(h_{\rm R}^{-1})^{\hat{\alpha}}_{\hat{\alpha}'}
   P^{\alpha'\hat{\alpha}'}(Z,\;(E^{\rm L}_{\beta}), \;
   (E^{\rm R}_{\hat{\beta}}))
\end{equation}
\paragraph     {Connections}
Connections are needed to define the kinetic terms for the ghost fields. 
A connection on the associated vector bundle $\widehat{M}\times_{\hat{H}} (S_L\otimes S_R)\stackrel{\pi}{\rightarrow}M$ is 
constructed from a connection on the principal bundle $\widehat{M}\stackrel{\pi}{\rightarrow} M$. We will now 
remind how this works. For any vector 
field $\xi\in \mbox{Vect}(M)$, a connection in the principal bundle defines a lift 
$\xi'\in \mbox{Vect}(\widehat{M})$, which is $\hat{H}$-invariant in the sense that for any $\chi\in \hat{\bf h}$ 
the corresponding vector field $v(\chi)$ commutes with $\xi'$:
\begin{equation}\label{AtiyahAlgebroid}
   [v(\chi),\xi'] = 0 \mbox{ \tt\small for any } \chi \in \hat{\bf h}
\end{equation}
For any representation $\rho\;:\; \hat{\bf h}\to \mbox{End}(V)$, sections of the associated bundle
$\widehat{M}\times_{\hat{H}} V\stackrel{\pi}{\rightarrow}M$ can be understood as maps $\sigma\;:\;\widehat{M}\rightarrow V$, invariant under $\hat{H}$ 
in the following sense:
\begin{equation}\label{SigmaIsInvariant}
   {\cal L}_{v(\chi)}\sigma = \rho(\chi)\sigma \;\mbox{ \tt\small for any } \chi \in \hat{\bf h}
\end{equation}
where ${\cal L}$ is the Lie derivative. Eq. (\ref{AtiyahAlgebroid}) implies that for any $\sigma$ satisfying 
(\ref{SigmaIsInvariant}), ${\cal L}_{\xi'}\sigma$ also satisfies (\ref{SigmaIsInvariant}). This means that the lift $\xi\mapsto\xi'$ consistently
defines the action of $\xi$ on the sections of the associated vector bundle. 

Let us explain how a connection in the principal bundle $\widehat{M}\to M$ defines a
kinetic term for the ghosts. Consider the ghost $\lambda_L$; in the flat space limit it is 
a left-moving field. In the general curved space, the kinetic term for $\lambda_L$ should
involve the derivative $\partial_-\lambda_L$. A point of the target space is 
$(Z,(E^{\rm L}_{\alpha}), (E^{\rm R}_{\hat{\alpha}}), \lambda_L,\lambda_R)$. The worldsheet is foliated by the characteristics. Let us
consider the right-moving characteristic $\tau^+=\mbox{const}$. It is parametrized by the 
$\tau^-$:
\begin{equation}
 (Z(\tau^-),\;(E^{\rm L}_{\alpha}(\tau^-)),\; (E^{\rm R}_{\hat{\alpha}}(\tau^-)), \;
 \lambda_L(\tau^-),\;\lambda_R(\tau^-))  
\end{equation}
Let us choose a representative so that 
$\left({dZ(\tau^-)\over d\tau^-},\;({dE^{\rm L}_{\alpha}(\tau^-)\over d\tau^-}),\; 
       ({dE^{\rm R}_{\hat{\alpha}}(\tau^-)\over d\tau^-})\right)$ is a horizonthal vector, in the sense defined by
the principal bundle connection in $\widehat{M}$. Then the kinetic term is:
\begin{equation}
   \int d\tau^+ d\tau^-\;\left(w^{\rm L}_+\;,\;{d\lambda_L\over d\tau^-}\right)
\end{equation}
where $w_+^{\rm L}$ is the conjugate momentum to $\lambda_L$.

\subsection{Lorentz superspace}\label{sec:LorentzSuperspace}
There is a way to canonically fix ${\bf R}_L^{\times}\times {\bf R}_R^{\times}$. In this paper we will use the 
variant of the formalism which has  ${\bf R}_L^{\times}\times {\bf R}_R^{\times}$ fixed. For us the gauge algebra 
is:
\begin{equation}\label{LorentzGaugeAlgebra}
   {\bf h} = {\bf h}_L \oplus {\bf h}_R = spin(1,9)_L\oplus spin(1,9)_R
\end{equation}
This version of the formalism is called ``Lorentz superspace''. We will now 
review how the Lorentz superspace is derived, as much as we understand. 

Consider the sigma-model (\ref{ActionOfBerkovitsAndHowe}) and let us {\bf integrate out} $d_{\alpha}$ and $\tilde{d}_{\hat{\beta}}$. It 
turns out that it is always possible to choose the gauge so that the coupling 
to the ghosts is only through the {\bf traceless} currents\footnote{N.~Berkovits, private communication; notice that we semiautomatically
arrived at this gauge in our study of linearized excitations of $AdS_5\times S^5$ in
\cite{Chandia:2013kja}.}:
\begin{equation}
   (w^{\rm L}_+\Gamma_{mn}\lambda_L) \mbox{ \tt\small and } (w^{\rm R}_-\Gamma_{mn}\lambda_R)
\end{equation}
The $u(1)$ combinations $(w^L_+\lambda_L)$ and $(w^R_-\lambda_R)$ appear only in the kinetic terms
$(w^L_+\partial_-\lambda_L)$ and $(w^R_-\partial_+\lambda_R)$. This fixes the gauge from $\hat{\bf h}_L\oplus \hat{\bf h}_R$ to ${\bf h}_L\oplus {\bf h}_R$. 
In the language of the present paper this simply means that we can use a slightly
simpler $\widehat{M}$. A point of this simplified  $\widehat{M}$ is a point $x\in M$ and a point in the
orbit of $H$ in ${\cal S}(x)\subset T_xM$; the simplification is in replacing the orbit of 
$\hat{H} = Spin(1,9)_L\times {\bf R}^{\times}_L\times Spin(1,9)_R\times {\bf R}^{\times}_R$ with the orbit of
$H = Spin(1,9)_L\times Spin(1,9)_R$. As in Section \ref{sec:FromQtoSUGRA}, we can still trade the BRST
operator for an $H$-invariant vector field on $\widehat{M}$. This statement is somewhat 
nontrivial, because what if the BRST operator $Q$ involves a rescaling of $\lambda_L$ and 
$\lambda_R$? Let us consider the action of $Q_R$ on $\lambda_L$:
\begin{equation}
   Q_R\lambda_L^{\alpha} = \lambda^{\hat{\alpha}}_RX_{\hat{\alpha}}{}^{\alpha}_{\beta}\lambda_L^{\beta}
\end{equation}
In particular, the $Q_R$ variation of the kinetic term $w^L_{\alpha+}\partial_-\lambda^{\alpha}_L$ gives the term
$w_{\alpha+}^LX_{\hat{\alpha}}{}^{\alpha}_{\beta}\lambda_L^{\beta}\partial_-\lambda_R^{\hat{\alpha}}$ which has nothing to cancel unless if $X_{\hat{\alpha}}$ is traceless, {\it i.e.}
if $X_{\hat{\alpha}}{}^{\alpha}_{\alpha}\neq 0$. (In this case it is cancelled by the variation of the connection on
which  $w^L_{\alpha+}\partial_-\lambda^{\alpha}_L$ depends, implicitly in our language.) Now consider the action
of $Q_L$ on $\lambda_L$:
\begin{equation}\label{QLLambdaL}
   Q_L\lambda_L^{\alpha} = \lambda_L^{\alpha}X_{\alpha}{}^{\beta}_{\gamma}\lambda_L^{\gamma}
\end{equation}
Now it is even meaningless to ask if $X_{\alpha}$ is traceless or not, because  $X_{\alpha}$ is
only defined by (\ref{QLLambdaL}) up to a shift transformation of Section \ref{sec:ShiftGaugeTransformations}. We therefore use
these shift transformations to remove the trace of $X_{\alpha}$. Then we have to remember that
when we work in the Lorentz superspace formalism, the shift transformations have 
their parameter restricted to:
\begin{equation}
\Gamma^n_{\alpha\bullet}h_L^{\bullet n } = 0
\end{equation}

\section{Worldsheet currents, quadratic-linear algebroid}\label{sec:Currents}
Let us consider the algebroid $\cal A$ over $\widehat{M}$ freely generated by ${\bf D}$ satisfying 
Eq. (\ref{SpecialD}) where ${\bf A}_m^L$ and ${\bf A}_m^R$ are free and  $R_{\alpha\beta}^{LL}$, $R_{\dot{\alpha}\dot{\beta}}^{RR}$ and $R_{\alpha\dot{\beta}}^{LR}$ are {\em same}
sections of $\Gamma(T\widehat{M}/M)$ as in Eq. (\ref{SpecialD}). The definition of ${\cal A}$ is a direct 
generalization of the definition of ${\cal L}_{\rm tot}$ in \cite{Mikhailov:2012uh}. We ``leave alone'' the 
vertical 
generators $R_{\alpha\beta}^{LL}$, $R_{\dot{\alpha}\dot{\beta}}^{RR}$ and $R_{\alpha\dot{\beta}}^{LR}$ in the sense that their commutation relations are 
postulated as the commutation relation in $\Gamma(T\widehat{M}/M)$. But we consider ${\bf D}^L_{\alpha}$ and ${\bf D}^R_{\hat{\alpha}}$
and  ${\bf A}_m^L$ and ${\bf A}_m^R$ as {\em free} generators modulo the relations (\ref{SpecialD}).

\paragraph     {(Open question:} Does ${\cal A}$ satisfy a PBW theorem?) 

\vspace{10pt}\noindent
From now on we will use letters ${\bf D}^L_{\alpha}$ and ${\bf D}^R_{\hat{\alpha}}$ to denote the generators of the
algebroid. The vector fields defined in Eq. (\ref{DefDComponents}) will now be interpreted as the
corresponding values of the anchor and therefore denoted $a({\bf D}^L_{\alpha})$ and $a({\bf D}^R_{\hat{\alpha}})$ 
(instead of simply ${\bf D}^L_{\alpha}$ and ${\bf D}^R_{\hat{\alpha}}$):
\begin{align}
   {\bf D}^L_{\alpha},\; {\bf D}^R_{\hat{\alpha}},\; {\bf A}^L_m,\; {\bf A}^R_m
   \;:&\; \mbox{ \small\tt generators of the algebroid } {\cal A}
   \nonumber \\    
   a({\bf D}^L_{\alpha}),\; a({\bf D}^R_{\hat{\alpha}}),\; 
   a({\bf A}^L_m),\; a({\bf A}^R_m)
   \;:&\; \mbox{ \tt\small vector fields on } \widehat{M}
   \nonumber
\end{align}
We introduce a bidirectional filtration on ${\cal A}$ in the following sense. For $n>0$, we 
will say that $\xi\in {\cal A}_{\leq n}$ if $\xi$ can be represented as a nested supercommutator
of $\leq n$ generators ${\bf D}^L_{\alpha}$. For $n<0$, we will say that $\xi\in {\cal A}_{\geq n}$ if $\xi$ can be 
represented as a nested supercommutator of $\leq |n|$ generators ${\bf D}^R_{\hat{\alpha}}$. Notice that
the expression containing nested supercommutators of both ${\bf D}^L_{\alpha}$ and ${\bf D}^R_{\hat{\alpha}}$ can be
reduced to expressions containing either all  ${\bf D}^L_{\alpha}$ or all ${\bf D}^R_{\hat{\alpha}}$.

\subsection{Basic consequences of the defining relations}\label{sec:BasicConsequences}
We observe that the basic commutation relations of (\ref{SpecialD}) imply the existence of ${\bf W}_L^{\alpha}$ 
such that:
\begin{align}
   [\;{\bf D}^L_{\alpha}\;,\;{\bf A}^L_m\;]\;=\;\Gamma_{m\alpha\beta} 
   {\bf W}_L^{\beta} \mbox{ mod } {\cal A}_{\leq 1}
\end{align}
Furthermore, notice the existence of ${\bf F}_{[mn]}$ such that:
\begin{align}
   \{\;{\bf D}^L_{\alpha}\;,\;{\bf W}_L^{\beta}\;\}\;=\;
   (\Gamma^{mn})_{\alpha}^{\beta} {\bf F}_{[mn]} 
   \mbox{ mod } {\cal A}_{\leq 2}
\end{align}
Indeed:
\begin{align}
   \Gamma_{m\beta(\gamma} \{{\bf D}^L_{\alpha)}\;,\;{\bf W}^{\beta}_L\}\;=\;
   \{{\bf D}^L_{(\alpha}\;,[{\bf D}^L_{\gamma)}\;,\;{\bf A}^L_m]\}\;=\;
   \Gamma_{\alpha\gamma}^n[{\bf A}^L_n\;,\;{\bf A}^L_m] \mbox{ mod } {\cal A}_{\leq 2}
\end{align}
and
\begin{equation}
   10 \{{\bf D}_{\alpha}^L\;,\;{\bf W}^{\alpha}_L\} \; = \;
   \Gamma_m^{\gamma\alpha}\Gamma^m_{\alpha\beta} 
   \{ {\bf D}^L_{\gamma}\;,\;{\bf W}^{\beta}_L\} \; = \;
   \Gamma_m^{\gamma\alpha}\{{\bf D}_{\gamma},[{\bf D}_{\alpha},{\bf A}_m]\}
   =0 \mbox{ mod } {\cal A}_{\leq 2}
\end{equation}
This implies the existence of ${\bf F}_{[mn]}$.

\subsection{Worldsheet currents}
Remember that the string worldsheet is spanned by the left-moving 
characteristics $\tau^-=\mbox{const}$. Consider an observer moving along a 
characteristic with the constant velocity $\dot{\tau}^+ = 1$. The velocity vector can be 
decomposed via the {\em worldsheet currents}:
\begin{align}\label{PartialPlusZ}
   \partial_+ Z^{\bf M} = 
   \widetilde{J}_{0+}^{L\bf M} + \widetilde{J}_{0+}^{R\bf M} + 
   \widetilde{J}_{+}^{\alpha} a^{\bf M}({\bf D}^L_{\alpha}) + 
   \widetilde{\Pi}_+^m a^{\bf M}({\bf A}^L_m) + 
   \widetilde{\psi}_{\alpha+}a^{\bf M}({\bf W}_L^{\alpha})
\end{align}
Here we used the abbreviation:
\begin{align}
   \widetilde{J}_{0+}^{L{\rm M}} = \widetilde{J}_{0+}^{L[mn]}a^{\rm M}(t^{L0}_{[mn]}) 
\end{align}
where $t^0_{[mn]}$ are generators of ${\bf h}_L$. 

Notice that the ``currents'' $\widetilde{J}_{0+}^{L[mn]}$, $\widetilde{I}^L_{0+}$, $\widetilde{J}^{\alpha}_+$, $\widetilde{\Pi}^m_+$, $\widetilde{\psi}_{\alpha+}$ are local functions 
on the phase space. We will denote the space of such functions $\mbox{Loc}({\cal X})$:
\begin{align}
   {\cal X} \;&=\; \mbox{\tt\small phase space }
   \nonumber\\      
   \mbox{Loc}({\cal X}) \;&=\; {
     \mbox{\tt\small the space of local functions on $\cal X$}
   }
   \nonumber
\end{align}
At the same time, $a(t^{L0})$, $a({\bf D}_{\alpha}^L)$, $a({\bf A}^L_m)$ and $a({\bf W}^{\alpha}_L)$ are vector fields on $\widehat{M}$.
Notice that a function $f(Z)$ on $\widehat{M}$ and a point $(\tau^+,\tau^-)$ on the worldsheet 
define a function on $\cal X$, namely $f(Z(\tau^+,\tau^-))$. In this sense, we should think 
of $\partial_+Z$ as an element of the space:
\begin{equation}
 {\cal V} =  \mbox{Loc}({\cal X})\;\otimes_{\mbox{Fun}(\widehat{M})}\; \mbox{Vect}(\widehat{M})
\end{equation}
This is {\em not} an algebroid over $\cal X$, because genarally speaking there is no 
way to lift a vector field on $\widehat{M}$ to a vector field on the phase space. But 
this is possible if the vector field generates a symmetry of the sigma-model.
When two elements $X\in {\cal V}$ and $Y\in {\cal V}$ both correspond to some symmetry of the
sigma-model, then it is possible to define the commutator $[X,Y]$. Another way 
of turning ${\cal V}$ into an algebroid is to go off-shell, {\it i.e.} replace the $\cal X$ 
with the space of off-shell configurations ${\cal X}_{\rm OS}$.

\subsection{Tautological Lax pair}
\subsubsection{The case of $AdS_5\times S^5$}
Consider the sigma model of the classical string in $AdS_5\times S^5$. It is 
classically integrable. There is a Lax pair, which depends on the spectral 
parameter $z$. At some
particular value of $z$, the Lax pair becomes tautological, the zero curvature 
equations being just the Maurer-Cartan equation for the worldsheet currents. We 
will now briefly review how this goes.

The current is $J= - dg g^{-1}$. For any representation of ${\bf g}$ with generators $t_a$,
it is straightforward to verify the Maurer-Cartan equation:
\begin{equation}\label{UsualLaxPair}
   \left[
      {\partial\over\partial\tau^+} + J_+^at_a \;,\;
      {\partial\over\partial\tau^-} + J_-^bt_b 
   \right] \;=\; 0
\end{equation}
We will need a slight variation of this construction. Let $\tilde{\bf g}$ be the Lie 
superalgebra obtained from $\bf g$ by changing the sign of the anticommutators (all
the commutators are the same, but all the anticommutators have the opposite 
sign). The left regular representation of $\tilde{\bf g}$ on the space of functions on the 
group manifold of $G$ is defined as follows:
\begin{equation}
   (L(\xi)f)(g) = \left.{d\over dt}\right|_{t=0}f(e^{-t\xi}g)
\end{equation}
This means that:
\begin{equation}\label{RegularRepresentation}
   {\partial f(g(\tau^+,\tau^-))\over\partial\tau^{\pm}} - (L(J)f)(g(\tau^+,\tau^-)) = 0
\end{equation}
We get\footnote{notice the difference in sign between (\ref{RegularRepresentation}) and (\ref{TautologicalLaxPairAdS})}:
\begin{equation}\label{TautologicalLaxPairAdS}
   \left[
      {\partial\over\partial\tau^+} + L(J_+) \;,\;
      {\partial\over\partial\tau^-} + L(J_-) 
   \right] \;=\; 0
\end{equation}
Eq. (\ref{TautologicalLaxPairAdS}) is almost the particular case of (\ref{UsualLaxPair}) corresponding to the left
regular representation. The only difference is that the left regular 
representation, as we defined it, is the representation of $\tilde{\bf g}$ and not $\bf g$. But at
the same time, notice that the odd-odd terms in $L(J_+)$ are of the form: 
$(-\partial_+\theta^{\alpha}+\ldots)\left({\partial\over\partial\theta^{\alpha}}
+\ldots\right)$ where $(-\partial_+\theta^{\alpha}+\ldots)$ is $J^{\alpha}_+$ and $\left({\partial\over\partial\theta^{\alpha}}
+\ldots\right) = \tilde{t}_{\alpha}$ is 
the corresponding generator of $\tilde{\bf g}$, let us call it $\tilde{t}_{\alpha}$. Notice that $\tilde{t}_{\alpha}$ 
{\em anti}-commutes with $J^{\alpha}_+$, while in (\ref{UsualLaxPair}) by definition $J^a$ commute with $t_a$.
In spite of this subtlety, the two definitions are actually equivalent. Given
a Lax pair in the sense of (\ref{TautologicalLaxPairAdS}), let us replace every term of the form 
$J^{\alpha}\tilde{t}_{\alpha}$ with $J^{\alpha}\tilde{t}_{\alpha}(-)^F$. Notice that $\tilde{t}_{\alpha}(-)^F$ are the generators of some 
representation of $\bf g$ (which should also be called ``left regular''), and also 
that $\tilde{t}_{\alpha}(-)^F$ commutes with $J^{\alpha}$. Therefore we obtained the Lax pair in the sense
of (\ref{UsualLaxPair}).

We can interpret the 
operator ${\partial\over\partial\tau^{\pm}} + L(J_{\pm})$ in the following way. Consider the space ${\cal X}_{\mbox{\tiny OS}}$ of all 
field configurations (off-shell) in the classical sigma-model. Let $\mbox{Loc}({\cal X}_{\mbox{\tiny OS}})$ 
denote the space of all local functions on ${\cal X}_{\mbox{\tiny OS}}$. Let us consider the space:
\begin{equation}\label{TensorProductOverFunM}
   \mbox{Fun}(\widehat{M})\;\otimes_{\mbox{Fun}(\widehat{M})} \;
 \mbox{Loc}({\cal X}_{\mbox{\tiny OS}})
\end{equation}
This is, obviously, the same as $\mbox{Loc}({\cal X}_{\mbox{\tiny OS}})$. Let us, however, define the action 
of the Lax operator on this space, as follows: ${\partial\over\partial \tau^{\pm}}$ acts only on $\mbox{Loc}({\cal X}_{\mbox{\tiny OS}})$ and 
$L(J_{\pm})$ acts only on $\mbox{Fun}(\widehat{M})$. Our point here is that this action is correctly 
defined. For example, the action of ${\partial\over\partial\tau^+} + \alpha L(J_+)$ with $\alpha\neq 1$ would not be 
correctly defined on (\ref{TensorProductOverFunM}), because it would act differently on $f\otimes \phi$ and 
$1\otimes f\phi$. 

\subsubsection{General case}
Consider the velocity of the coordinate function $Z^{\widehat{M}}$ pulled back on the string
worldsheet:
\begin{align}
   {\partial Z^{\widehat{M}}(\tau^+,\tau^-)\over\partial\tau^+} \;= &\;
   \widetilde{J}_{0+}^{{\rm L}[mn]} a^{\widehat{M}}(t^{{\rm L}0}_{[mn]}) 
   + \widetilde{J}_{0+}^{{\rm R}[mn]} a^{\widehat{M}}(t^{{\rm R}0}_{[mn]}) \;+
   \nonumber \\    
   &\; 
   + \widetilde{J}_{+}^{\alpha} a^{\widehat{M}}({\bf D}^L_{\alpha}) 
   + \widetilde{\Pi}_+^m a^{\widehat{M}}({\bf A}^L_m) 
   + \widetilde{\psi}_{\alpha+}a^{\widehat{M}}({\bf W}_L^{\alpha})
   \label{DZDtauPlus}
\end{align}
We write $Z^{\widehat{M}}$ instead of simply $Z^M$, to stress that the coordinates include also
the fiber. In the $AdS_5\times S^5$ language, $Z^{\widehat{M}}$ would parametrize $PSU(2,2|4)$ rather 
than AdS. The  terms $\widetilde{J}_{0+}^{{\rm L}[mn]} a^{\widehat{M}}(t^{{\rm L}0}_{[mn]})$ and $\widetilde{J}_{0+}^{{\rm R}[mn]} a^{\widehat{M}}(t^{{\rm R}0}_{[mn]})$ are vertical (along the
fiber). 

Then $\left[{\partial\over\partial\tau^+}\;,{\partial\over\partial\tau^-}\right]Z^{\widehat{M}} = 0$ leads to the tautological zero curvature equation:
\begin{align}
   &
   {\partial\over\partial\tau^+} \left(
      \widetilde{J}^{L{\rm M}}_{0-} + \widetilde{J}^{R{\rm M}}_{0-} +
      \widetilde{J}_{-}^{\alpha} a({\bf D}^L_{\alpha}) + 
      \widetilde{\Pi}_-^m a({\bf A}^L_m) + 
      \widetilde{\psi}_{\alpha-}a({\bf W}_L^{\alpha})
   \right)\;-
   \nonumber\\   
   -\; & 
   {\partial\over\partial\tau^-} \left(
      \widetilde{J}^{L{\rm M}}_{0+} + \widetilde{J}^{R{\rm M}}_{0+} +
      \widetilde{J}_{+}^{\alpha} a({\bf D}^L_{\alpha}) + 
      \widetilde{\Pi}_+^m a({\bf A}^L_m) + 
      \widetilde{\psi}_{\alpha+}a({\bf W}_L^{\alpha})
   \right)\;+
   \nonumber\\   
   +\; & 
   \Big[
         \widetilde{J}^{L{\rm M}}_{0+} + \widetilde{J}^{R{\rm M}}_{0+} +
         \widetilde{J}_{+}^{\alpha} a({\bf D}^L_{\alpha}) + 
         \widetilde{\Pi}_+^m a({\bf A}^L_m) + 
         \widetilde{\psi}_{\alpha+}a({\bf W}_L^{\alpha})
         \;,
      \nonumber\\   
      &  
      \;\;
         \widetilde{J}^{L{\rm M}}_{0-} + \widetilde{J}^{R{\rm M}}_{0-} +
         \widetilde{J}_{-}^{\alpha} a({\bf D}^L_{\alpha}) + 
         \widetilde{\Pi}_-^m a({\bf A}^L_m) + 
         \widetilde{\psi}_{\alpha-}a({\bf W}_L^{\alpha})
      \Big]\;=0
\end{align}
In this formula $\partial\over\partial\tau^+$ in the first line and $\partial\over\partial\tau^-$ in the second line only act on
the currents $\widetilde{J},\widetilde{\Pi},\widetilde{\psi}$ and do not act on $a({\bf D}), a({\bf A}), a({\bf W})$. The commutator is 
the commutator of the vector fields, {\it e.g.}:
\begin{equation}
   \left[
      \widetilde{\Pi}_+^m a({\bf A}^L_m) \;,\;
      \widetilde{\Pi}_-^n a({\bf A}^L_n) 
   \right]\; =\;
   \widetilde{\Pi}_+^m \widetilde{\Pi}_-^n 
   \left[
      a({\bf A}^L_m) \;,\;a({\bf A}^L_n) 
   \right]
\end{equation}

\paragraph     {Consequences of $[Q_L,\partial_+]=0$} 
Let us consider the BRST variation:
\begin{equation}\label{QLZ}
   \epsilon Q_L Z^{\widehat{M}} = \epsilon \lambda_L^{\alpha} a^{\widehat{M}}({\bf D}^L_{\alpha})
\end{equation}
We have two vector fields on the phase space, $Q_L$ and ${\partial\over\partial\tau^+}$. They commute:
\begin{align}
   &
   (\epsilon Q_L\widetilde{J}_{0+}^{L[mn]})a(t^L_{[mn]}) \;+\; 
   (\epsilon Q_L\widetilde{J}_{0+}^{R[mn]})a(t^R_{[mn]}) \;+\;
   \nonumber \\ 
   +\;& 
   (\epsilon Q_L\widetilde{J}_+^{\alpha})a({\bf D}^L_{\alpha})\;+\;
   (\epsilon Q_L\widetilde{\Pi}_+^m)a({\bf A}^L_m)\;+\;
   (\epsilon Q_L\widetilde{\psi}_{\alpha+})a({\bf W}_L^{\alpha})\;-\;
   \nonumber \\   
   -\;&  
   \partial_+(\epsilon\lambda_L^{\beta})a({\bf D}_{\beta}^L) 
   +\;
   \nonumber \\   
   +\;&
   \widetilde{J}_{0+}^{L[mn]}a\left(
      \epsilon\lambda_L^{\beta}[{\bf D}_{\beta}^L ,\;t^0_{L[mn]}]+
   \right) \;+\;
   \widetilde{J}_{0+}^{R[mn]}a\left(
      \epsilon\lambda_L^{\beta}[{\bf D}_{\beta}^L ,\;t^0_{R[mn]}]
   \right) +
   \nonumber \\  
   +\;&  
   \widetilde{J}_+^{\alpha}a\left(
      \epsilon\lambda_L^{\beta}\{{\bf D}_{\beta}^L\;,\;{\bf D}^L_{\alpha}\} 
   \right) \;+\;
   \nonumber \\  
   +\;&
   \widetilde{\Pi}_+^ma\left(
      \epsilon\lambda_L^{\beta}[{\bf D}_{\beta}^L,\;{\bf A}^L_m] 
   \right) \;+\;
   \nonumber \\  
   +\;& 
   \widetilde{\psi}_{\alpha+}a\left(
      \epsilon\lambda_L^{\beta}\{{\bf D}_{\beta}^L,\;{\bf W}^{\alpha}_L\} 
   \right)\;\;=\; 0 
   \label{CommutatorOfQLWithDPlus}
\end{align}
In particular, we can say something about $Q_L\widetilde{\psi}_{\alpha+}$. Let us define the superfield
$C^{\alpha}_{\beta\gamma}(Z)$ by the following formula:
\begin{equation}\label{DefCAlphaBetaGamma}
   \{\; a({\bf D}^L_{\beta}) \;,\; a({\bf W}^{\alpha}_L) \;\}\;=\;
   C^{\alpha}_{\beta}{}_{\gamma}\;a({\bf W}^{\gamma}_L)  \;
   \mbox{\tt\small mod } {\cal A}_{[0,2]}
\end{equation}
where $\mbox{\small\tt mod }{\cal A}_{[0,2]}$ stands for a linear combination of $a({\bf D}_{\alpha}^L)$ , $a({\bf A}_m^L)$, $a(t^L_{[mn]})$ 
and $a(t^R_{[mn]})$. From (\ref{CommutatorOfQLWithDPlus}) we read:
\begin{equation}
   (\epsilon Q_L\widetilde{\psi}_{\alpha+}) a\left({\bf W}_L^{\alpha}\right) +
   a\left(
      [\epsilon \lambda_L^{\gamma}{\bf D}^L_{\gamma}\;,\;
      \widetilde{\psi}_{\alpha+}{\bf W}_L^{\alpha}]
   \right) +
   a\left(
      [\epsilon\lambda_L^{\gamma}{\bf D}^L_{\gamma}\;,\;
      \widetilde{\Pi}^m_+{\bf A}_L^m]
   \right)\;\in\;{\cal A}_{[0,2]}
\end{equation}
and therefore:
\begin{equation}\label{QLofPsi}
   Q_L\widetilde{\psi}_{\alpha+} - (\widetilde{\psi}_+ C_{\alpha} \lambda_L)
   + \lambda_L^{\gamma}\widetilde{\Pi}^m_+ \Gamma_{\gamma\alpha}^m = 0
\end{equation}
Similarly we have:
\begin{align}
   Q_L \widetilde{\Pi}^m_+ + 
   \widetilde{J}^{\alpha}_+ \Gamma^m_{\alpha\beta} \lambda_L^{\beta} + 
   \widetilde{\psi}_{\alpha+}F^{\alpha}_{\beta}{}^m\lambda_L^{\beta} = 0
\end{align}
with some $F^{\alpha m}_{\beta}$ originating from $[\epsilon \lambda_L^{\gamma}{\bf D}^L_{\gamma}\;,\;
      \widetilde{\psi}_{\alpha+}{\bf W}_L^{\alpha}]$. The nilpotence of $Q_L$ 
implies:
\begin{align}
   Q_L^2\widetilde{\psi}_{\alpha+} \;=&\;
   - (\widetilde{\psi}_+(Q_LC_{\alpha})\lambda_L) +
   ((\widetilde{\psi}_+C\lambda_L)C_{\alpha}\lambda_L) - 
   (\lambda_L Q_L(\widetilde{\Pi}^m_+)\Gamma^m)_{\alpha}\;=
   \nonumber  \\   
   \;=&\;
   - R_{\alpha_1\alpha_2}{}^{\alpha'}_{\alpha}\widetilde{\psi}_{\alpha'+}
\end{align}
Happily, the only part of $Q_L\widetilde{\Pi}^m_+$ which gives a nonvanishing contribution is
proportional to $\widetilde{\psi}_{\alpha+}$; let us extract its coefficient:
\begin{align}
   &\;   \lambda^{\alpha_1}_L\lambda^{\alpha_2}_L
   a({\bf D}^L_{\alpha_1})C^{\beta}_{\alpha_2\alpha} = 
   \nonumber \\ 
   = &\; \lambda_L^{\alpha_1}\lambda_L^{\alpha_2}
   C^{\beta}_{\alpha_1\delta}C^{\delta}_{\alpha_2\alpha} 
   + \lambda_L^{\alpha_1}\lambda_L^{\alpha_2} R_{\alpha_1\alpha_2}{}^{\beta}_{\alpha} + 
   \lambda_L^{\alpha_1}\lambda_L^{\alpha_2}F^{\beta}_{\alpha_1}{}^m\Gamma^m_{\alpha_2\alpha}
   \label{QCLambda}
\end{align}

\subsection{Identification of $\widetilde{\psi}_{\alpha+}$}
We will now show that $\widetilde{\psi}_{\alpha+}$ can be identified as the matter part of the BRST
charge density.

Remember that $w^L_+$ is the momentum conjugate to $\lambda_L$ --- see Eq. (\ref{ActionOfBerkovitsAndHowe}).
Let us define\footnote{One could define $d_{\alpha+}$ through the density
of the BRST charge $Q_L$, which is $\lambda_L^{\alpha}d_{\alpha+}$. Such a 
definition
would only specify $d_{\alpha+}$ up to an addition of the terms of the form
$X^m \Gamma^m_{\alpha\beta}\lambda_L^{\beta}$ and $X^{klm}\Gamma^{klm}_{\alpha\beta}\lambda_L^{\beta}$. It is possible to reduce this ambiguity by defining $d_{\alpha+}$ from Eq. (\ref{DefDAlphaPlus}); this leaves only the ambiguity of the form
$X^m \Gamma^m_{\alpha\beta}\lambda_L^{\beta}$. In this sense, $d_{\alpha+}$ is ``better-defined'' than one might think.} $\psi_{\alpha+}$ and $d_{\alpha+}$ as follows:
\begin{align}
   \psi_{\alpha+}\;&= \left(
      \widetilde{\psi}_{\alpha+} - 
      w_{\circ+}^LC^{\circ}_{\bullet}{}_{\alpha} \lambda_L^{\bullet}
   \right)
   \label{PsiVsPsiTilde}\\
   d_{\alpha+}\;& = Q_L(w_{\alpha+}^L)
   \label{DefDAlphaPlus}
\end{align}
It follows:
\begin{equation}\label{QLdPlus}
   Q_Ld_{\alpha+} = -\lambda_L^{\alpha_1}\lambda_L^{\alpha_2}R_{\alpha_1\alpha_2}{}^{\beta}_{\alpha}w^{L}_{\beta+}\mbox{ \tt\small mod } (\_)_{m+}\Gamma^m_{\alpha\gamma}\lambda_L^{\gamma}
\end{equation}
Here ``$\mbox{ \small\tt mod }(\_)_{m+}\Gamma^m_{\alpha\gamma}\lambda_L^{\gamma}$'' means ``up to adding $u_{m+}\Gamma^m_{\alpha\gamma}\lambda_L^{\gamma}$ 
with some arbitrary $u_{m+}$''. Notice that $\lambda_L^{\alpha}d_{\alpha+}$ is the left BRST current.
This follows from the fundamental property of the formalism: the $U(1)_L$ 
charge of the left BRST current is $+1$. We have: 
\begin{align}
   Q_L\psi_{\alpha+} \;=&\; Q_L\widetilde{\psi}_{\alpha+} 
   - (d_{\circ+}C^{\circ}_{\bullet\alpha}\lambda^{\bullet}_L) - 
   (w_{\circ+} (QC^{\circ}_{\bullet\alpha}) \lambda^{\bullet}_L) \;=
   \nonumber \\  
   \;=&\;  
   (\widetilde{\psi}_{\triangleright+}C^{\triangleright}_{\bullet\alpha}\lambda^{\bullet}_L) - 
   \lambda_L^{\bullet}\widetilde{\Pi}^m_+\Gamma^m_{\bullet\alpha} - 
   (d_{\triangleright+}C^{\triangleright}_{\bullet\alpha}\lambda^{\bullet}_L) - 
   (w_{\triangleright+} (QC^{\triangleright}_{\bullet\alpha}) \lambda^{\bullet}_L) \;=
   \nonumber \\  
   =&\;
   ((w_{\triangleleft+}C^{\triangleleft}_{\circ\bullet}\lambda^{\circ}_L)
   C^{\bullet}_{\triangleright\alpha}\lambda^{\triangleright}_L) - 
   (w_{\triangleleft+} (QC^{\triangleleft}_{\bullet\alpha}) \lambda^{\bullet}_L) - \;
   \nonumber \\   
   & - \lambda_L^{\bullet}\widetilde{\Pi}^m_+\Gamma^m_{\bullet\alpha}  +
   ((\psi_+-d_+)C_\alpha\lambda_L)
\end{align}
It follows from Eq. (\ref{QCLambda}) that:
\begin{equation}
      ((w_+C\lambda_L)C_{\alpha}\lambda_L) - (w (QC_{\alpha}) \lambda_L) =
      - w_{\beta+}\lambda_L^{\alpha_1}\lambda_L^{\alpha_2} R_{\alpha_1\alpha_2}{}^{\beta}_{\alpha}
      \mbox{ \small\tt mod }(\_)_{m+}\Gamma^m_{\alpha\gamma}\lambda_L^{\gamma}
\end{equation}
Therefore:
\begin{equation}\label{QLPsi}
   Q_L \psi_{\alpha+} = 
   - \lambda_L^{\alpha_1}\lambda_L^{\alpha_2}R_{\alpha_1\alpha_2}{}^{\beta}_{\alpha} w^L_{\beta+} 
   +((\psi_{\circ+}-d_{\circ+})C^{\circ}_{\bullet\alpha}\lambda_L^{\bullet})
   \mbox{ \small\tt mod }(\_)_{m+}\Gamma^m_{\alpha\gamma}\lambda_L^{\gamma}
\end{equation}
Let us denote $\zeta_{\alpha+} = \psi_{\alpha+} - d_{\alpha+} \mbox{ \tt\small mod } (\_)_m\Gamma^m_{\alpha\gamma}\lambda_L^{\gamma}$. 

\refstepcounter{Theorems}
\noindent{\bf Theorem \arabic{Theorems}}: 
\begin{equation}\label{ZetaIsZero}
\zeta_{\alpha+}=0 \mbox{ \tt\small mod }(\_)_{m+}\Gamma^m_{\alpha\bullet}\lambda_L^{\bullet}
\end{equation}
\noindent{\bf Proof}: Comparing Eqs. (\ref{QLPsi}) and (\ref{QLdPlus}) we get: 
\begin{equation}\label{EquationOnZeta}
   Q_L\zeta_{\alpha+} = \zeta_{\circ+}C^{\circ}_{\bullet}{}_{\alpha}\lambda_L^{\bullet}\mbox{ \tt\small mod } (\_)_{m+}\Gamma^m_{\alpha\bullet}\lambda_L^{\bullet}
\end{equation}
(the same equation as (\ref{QLofPsi})).

It follows from the analysis of Eq. (\ref{EquationOnZeta}) in the flat space 
limit that any $\zeta_{\alpha+}$ satisfying (\ref{EquationOnZeta}) is of the form:
\begin{equation}
   \zeta_{\alpha+} = \phi\left( d_{\alpha_+} +  C^{\circ}_{\bullet}{}_{\alpha}w_{\circ+}\lambda_L^{\bullet}\right) + 
   B^{\circ}_{\bullet}{}_{\alpha}w_{\circ+}\lambda_L^{\bullet}
\end{equation}
where $\phi = \phi(Z)$  and $B^{\beta}_{\gamma\alpha} = B^{\beta}_{\gamma\alpha}(Z)$ are some functions.
Indeed, in the flat space limit, in the neighborhood of {\em any} point of $M$, 
if $\theta$ and $\lambda$ scale as $R^{-1/2}$ and $x$ as $R^{-1}$ and $w_{\pm}$ as $R^{-3/2}$, then $\zeta_{\alpha+}$ should 
be of the order $R^{-3/2}$; this means that the coefficients of $\widetilde{J}_+$ and $\widetilde{\Pi}_+$ in $\zeta_{\alpha+}$ 
are zero. The leading term in the flat space expansion of $\zeta_{\alpha+}$ is then of the 
form $\phi_{\alpha}^{\beta}d_{\beta+}$, and its BRST variation is in the leading order 
$(Q_L\phi_{\alpha}^{\beta})d_{\beta+} + \phi_{\alpha}^{\beta}\Gamma^m_{\beta\gamma}\Pi_+^m\lambda_L^{\gamma}$. Therefore the vanishing of the leading term in 
Eq. (\ref{EquationOnZeta}) up to $(\_)_{m+}\Gamma^m_{\alpha\bullet}\lambda^{\bullet}$ implies that $\phi_{\alpha}^{\beta}$ is proportional to $\delta_{\alpha}^{\beta}$.

Notice that Eq. (\ref{EquationOnZeta}) is satisfied when $\phi=\mbox{const}$ and $B^{\beta}_{\gamma\alpha} = 0$. When $\phi$ is not 
constant, the vanishing of the coefficient of $d_+$ in (\ref{EquationOnZeta}) implies:
\begin{equation}\label{BViaPhi}
   B^{\beta}_{\gamma\alpha} \;=\; 
   \delta^{\beta}_{\alpha} D^L_{\gamma}\phi - 
   {1\over 2}\Gamma^m_{\alpha\gamma} \Gamma_m^{\beta\bullet}D^L_{\bullet}\phi 
\end{equation}
and the vanishing of the coefficient of $w_+\lambda_L\lambda_L$ implies:
\begin{equation}
   \lambda_L^{\bullet}\lambda_L^{\bullet}\left(
      D^L_{\bullet}B^{\beta}_{\bullet\alpha} + 
      C_{\bullet\alpha}^{\circ}B^{\beta}_{\bullet\circ}
   \right)w_{\beta+} = (\_)_m \Gamma^m_{\alpha\bullet}\lambda_L^{\bullet}
\end{equation}
This is equivalent to the following equation being satisfied for any pure spinor $\lambda_L$:
\begin{equation}
     \lambda_L^{\bullet}\lambda_L^{\bullet}\lambda_L^{\bullet}\left(
      D^L_{\bullet}B^{\beta}_{\bullet\bullet} + 
      C_{\bullet\bullet}^{\circ}B^{\beta}_{\bullet\circ}
   \right) = 0
\end{equation}
Substitution of (\ref{BViaPhi}) gives:
\begin{equation}
   C^{\beta}_{\bullet\bullet}\lambda_L^{\bullet}\lambda_L^{\bullet} \lambda_L^{\bullet}
   D^L_{\bullet}\phi = 0
\end{equation}
This implies that either $C^{\beta}_{\bullet\bullet}\lambda_L^{\bullet}\lambda_L^{\bullet} =0$ for any $\lambda_L$, which is generally speaking not
the case, or $\lambda_L^{\bullet} D^L_{\bullet}\phi = 0$, which implies implies that $\phi=\mbox{const}$. In the case of 
$AdS_5\times S^5$ we know $\zeta_{\alpha+}$ is zero. Therefore $\phi=0$ and Eq. (\ref{ZetaIsZero}) follows.

This means that in terms of the sigma-model (\ref{ActionOfBerkovitsAndHowe}):
\begin{align}
   \widetilde{\psi}_{\alpha+} \;=&\; 
   P^{-1}_{\alpha\hat{\alpha}} E^{\hat{\alpha}}_M\partial_+Z^M
   \\     
   d_{\alpha+} \;=&\; \widetilde{\psi}_{\alpha+} -
    C_{\beta\alpha}^{\gamma}\lambda^{\beta}w_{\gamma+}
   \mbox{ \tt\small mod } (\_)_{m+}\Gamma^m_{\alpha\beta}\lambda_L^{\beta}
   \label{dInSUGRA}
\end{align}

\subsection{Identification of $C^{\beta}_{\alpha\gamma}$}
Let us consider the SUGRA superfields  $C^{\beta\hat{\alpha}}_{\gamma}$ and $P_{\alpha\hat{\alpha}}$ defined in Eq. (\ref{ActionOfBerkovitsAndHowe}).
(Notice that we use the same letter $C$ as for $C^{\beta}_{\alpha\gamma}$, but with a different set of 
indices; we hope this will not lead to confusion.)
Eq. (\ref{dInSUGRA}) implies that:
\begin{equation}\label{SigmaVsAlg}
   C^{\alpha}_{\beta}{}^{\hat{\gamma}} P^{-1}_{\hat{\gamma}\gamma} = -C^{\alpha}_{\beta\gamma}
\end{equation}
In particular, this implies that in the Lorentz superspace formalism (\ref{LorentzGaugeAlgebra}):
\begin{equation}\label{TracelessC}
C^{\alpha}_{\alpha}{}^{\hat{\gamma}} = 0
\end{equation} 
(This is not stated in \cite{Berkovits:2001ue}.) One difference of our approach with \cite{Berkovits:2001ue} is that we do 
not require that $T^{\alpha}_{\beta\gamma}=0$. In fact, it is difficult to define  $T^{\alpha}_{\beta\gamma}$ in our language.

We will now confirm this by comparing the ``shift'' gauge transformations 
defined in Eq. (61) of \cite{Berkovits:2001ue}. They correspond to the following variation of $\nabla^L_{\alpha}$:
\begin{align}
   \delta_h \nabla_{\alpha}^L  \;&= \omega_{\alpha}
   \\   
   \omega_{\alpha}{}^{\beta}_{\gamma} \;&= (\Gamma^k_{\alpha\bullet} h^{\bullet n})
   (\Gamma_n^{\beta\circ}\Gamma^k_{\circ\gamma})
\end{align}
where $h^{\alpha n}$ is a gauge parameter. In the Lorentz superspace formalism (\ref{LorentzGaugeAlgebra}) the 
shift parameter satisfies:
\begin{equation}\label{ShiftParameterIsSpinThreeHalves}
   \Gamma_{\alpha\beta}^nh^{\beta n} = 0
\end{equation} 
Let us determine the transformation of $\psi_{\alpha+}$ 
and $C^{\alpha}_{\beta\gamma}$.
\begin{align}
   \{\delta_h\nabla^L_{(\alpha_1},\nabla^L_{\alpha_2)}\} \;&=
   - (\Gamma^k_{(\alpha_1|\bullet} h^{\bullet n})\nabla^L_{\circ}
   (\Gamma_n^{\circ\bullet}\Gamma^k_{\bullet|\alpha_2)}) \;=
   \\   
   \;&= {1\over 2} \Gamma^p_{\alpha_1\alpha_2} h^{n\triangleright}
   \Gamma^p_{\triangleright\bullet} \Gamma_n^{\bullet\alpha}\nabla^L_{\alpha}
\end{align}
In other words:
\begin{align}
   \delta_h {\bf A}^p \;&=  h^{n\triangleright}\Gamma^p_{\triangleright\bullet}
   \Gamma_n^{\bullet\circ}\nabla^L_{\circ}
   \\  
   [\nabla_{\alpha}\;,\;\delta_h{\bf A}^p] \;&= 
   - h^{n\triangleright}\Gamma^p_{\triangleright\bullet}
   \Gamma_n^{\bullet\circ}\Gamma^k_{\circ\alpha}{\bf A}_k\;=
   \nonumber \\   
   \;&= 
   h^{n\triangleright}\Gamma^n_{\triangleright\bullet}\Gamma_p^{\bullet\circ}\Gamma^k_{\circ\alpha}{\bf A}_k - 
   2h^{p\bullet}\Gamma^k_{\bullet\alpha}{\bf A}_k\;=
   \nonumber \\   
   \;&= 
   -h^{n\triangleright}\Gamma^n_{\triangleright\bullet}\Gamma_k^{\bullet\circ}\Gamma^p_{\circ\alpha}{\bf A}_k - 
   2h^{p\bullet}\Gamma^k_{\bullet\alpha}{\bf A}_k + 
   2h^{n\bullet}\Gamma^n_{\bullet\alpha}{\bf A}_p
\end{align}
\begin{align}
   [\delta_h\nabla_{\alpha}\;,\;{\bf A}^p]\Gamma^p_{\alpha_1\alpha_2} \;&= 
   - 2 \omega_{\alpha}{}_{(\alpha_1}^{\bullet} {\bf A}^p\Gamma^p_{\alpha_2)\bullet} \;=
   \\   
   \;&= 
   - 2 (\Gamma^k_{\alpha\circ} h^{\circ n})
   (\Gamma_n^{\bullet\triangleleft}\Gamma^k_{\triangleleft(\alpha_1})  
   {\bf A}^p\Gamma^p_{\alpha_2)\bullet} \;=
   \\
   \;&=
   2(\Gamma^k_{\alpha\circ} h^{\circ n})\Gamma^n_{\alpha_1\alpha_2}{\bf A}^k 
   - 4 (\Gamma^k_{\alpha\circ} h^{\circ (k})\Gamma^{p)}_{\alpha_1\alpha_2}{\bf A}^p
\end{align}
where we used the gamma-matrix identity:
\begin{equation}
   \Gamma^p_{(\alpha_1|\bullet}\Gamma^{\bullet\triangleleft}_n\Gamma^k_{\triangleleft|\alpha_2)}=
   (\Gamma^p\Gamma_n\Gamma^k)_{(\alpha_1\alpha_2)} = (\Gamma^{(p}\Gamma_n\Gamma^{k)})_{(\alpha_1\alpha_2)} = - \delta^{pk}\Gamma^n_{\alpha_1\alpha_2} + 2\delta^{n(k}\Gamma^{p)}_{\alpha_1\alpha_2}
\end{equation}
and therefore:
\begin{equation}
   [\delta_h\nabla_{\alpha}\;,\;{\bf A}^p]\;= 2 \Gamma^k_{\alpha\circ}h^{\circ p}{\bf A}^k
   - 2(\Gamma^k_{\alpha\circ}h^{\circ k}){\bf A}^p 
   - 2(\Gamma^p_{\alpha\circ}h^{\circ k}) {\bf A}^k
\end{equation}
This implies: 
\begin{equation}
   \delta_h {\bf W}^{\alpha}_L = - h^{n\bullet}\Gamma^n_{\bullet\circ} \Gamma_k^{\circ\alpha}{\bf A}_k - 2 h^{\alpha k }{\bf A}^k
\end{equation}
\begin{align}
   \{\nabla^L_{\beta},\delta_h{\bf W}_L^{\alpha}\} \;&= 
   h^{n\bullet}\Gamma^n_{\bullet\circ} \Gamma_k^{\circ\alpha}[\nabla^L_{\beta},{\bf A}_k]
   + 2 h^{\alpha k} [\nabla^L_{\beta}, {\bf A}^k]
   \;= 
   \\   
   &=h^{n\bullet}\Gamma^n_{\bullet\circ} \Gamma_k^{\circ\alpha}\Gamma^k_{\beta\gamma}
   {\bf W}_L^{\gamma} + 2 h^{\alpha k} \Gamma_{\beta\gamma}^k {\bf W}_L^{\gamma}
\end{align}
At the same time:
\begin{align}
   \{\delta_h \nabla_{\beta}^L, {\bf W}_L^{\alpha}\} \;&=
   (h^{n\circ}\Gamma^k_{\circ\beta}) (\Gamma^{\alpha\bullet}_n\Gamma^k_{\bullet\gamma}) 
   {\bf W}_L^{\gamma} - 4 h^{k\circ}\Gamma_{\circ\beta}^k {\bf W}_L^{\alpha}
\end{align}
where the term $- 4 h^{k\circ}\Gamma_{\circ\beta}^k {\bf W}_L^{\alpha}$ corresponds to the trace part of $\omega$.
Therefore:
\begin{align}
   \delta_h \{\nabla_{\beta}^L,{\bf W}_L^{\alpha}\} \;=&\;
   h^{n\bullet}\Gamma^n_{\bullet\circ} \Gamma_k^{\circ\alpha}\Gamma^k_{\beta\gamma}
   {\bf W}_L^{\gamma} + 
   2 h^{\alpha k} \Gamma_{\beta\gamma}^k {\bf W}_L^{\gamma} \;+
   \nonumber \\ 
   &
   + (h^{n\bullet}\Gamma^k_{\bullet\beta}) (\Gamma^{\alpha\circ}_n\Gamma^k_{\circ\gamma}) 
   {\bf W}_L^{\gamma} - 4 h^{k\circ}\Gamma_{\circ\beta}^k {\bf W}_L^{\alpha}
\end{align}
This implies that\footnote{as a consistency check,  $\delta_h C^{\alpha}_{\alpha\gamma}=0$ and $\delta (\Gamma^{klmn})^{\beta}_{\alpha}C^{\alpha}_{\beta\gamma}=0$.}:
\begin{align}
   \delta_h C^{\alpha}_{\beta\gamma} =\;& 
   h^{n\bullet}\Gamma^n_{\bullet\circ} \Gamma_k^{\circ\alpha}\Gamma^k_{\beta\gamma} +
   2 h^{n\alpha} \Gamma_{\beta\gamma}^n  \;+
   \nonumber \\ 
   &
   + h^{n\bullet}\Gamma^k_{\bullet\beta} \Gamma^{\alpha\circ}_n\Gamma^k_{\circ\gamma}
   - 4 h^{n\bullet}\Gamma_{\bullet\beta}^n \delta^{\alpha}_{\gamma}
\end{align}
Therefore:
\begin{align}
   \delta_h C^{\alpha}_{\beta\gamma} + \delta_h(C^{\alpha}_{\beta}{}^{\hat{\alpha}} P^{-1}_{\hat{\alpha}\gamma})\;=&\;
   h^{n\bullet}\Gamma^n_{\bullet\circ} \Gamma_k^{\circ\alpha}\Gamma^k_{\beta\gamma} +
   2 h^{n\alpha} \Gamma_{\beta\gamma}^n  \;+
   \nonumber \\ 
   &
   + 2 h^{n\bullet}\Gamma^k_{\bullet(\beta|} \Gamma^{\alpha\circ}_n\Gamma^k_{\circ|\gamma)}
   - 4 h^{n\bullet}\Gamma_{\bullet\beta}^n \delta^{\alpha}_{\gamma}\;=
   \nonumber \\   
   =& - 4 h^{n\bullet}\Gamma_{\bullet\beta}^n \delta^{\alpha}_{\gamma}\;
   \mbox{ \tt\small mod } 
   (\_)_m\Gamma^m_{\alpha\beta}\lambda_L^{\beta}
\end{align}
Given (\ref{ShiftParameterIsSpinThreeHalves}), this is in agreement with (\ref{SigmaVsAlg}).   

\subsection{Ramond-Ramond fields}
The Ramond-Ramond bispinor $P^{\alpha\hat{\alpha}}$ must have a similar interpretations. 
Let us expand ${\bf D}^R_{\hat{\alpha}}$ in terms of ${\bf W}_L^{\alpha}, {\bf A}^m_L, {\bf D}^L_{\alpha}$ and $T\widehat{M}/M$. We will get:
\begin{equation}
   {\bf D}^R_{\hat{\alpha}}  = P_{\hat{\alpha}\alpha}{\bf W}_L^{\alpha} + \ldots
\end{equation}
where $\ldots$ stand for the terms proportional to ${\bf A}_L^m$ and ${\bf D}^L_{\alpha}$  and $T\widehat{M}/M$.
(Again, the coefficients of  ${\bf A}_L^m$ and ${\bf D}^L_{\alpha}$ are not defined unambiguously, but
the coefficient of ${\bf W}_L^{\alpha}$ is well-defined.) We conjecture that $P_{\hat{\alpha}\alpha}$ is the 
inverse of the $P^{\alpha\hat{\alpha}}$ of (\ref{ActionOfBerkovitsAndHowe}), however we do not have a proof. Similarly:
\begin{equation}
   {\bf D}^L_{\alpha}  = P_{\alpha\hat{\alpha}}{\bf W}_R^{\hat{\alpha}} + \ldots
\end{equation}
This is probably the most concise definition of the Ramond-Ramond bispinor
in the framework of the pure spinor sigma-model.

\subsection{Weighing anchor}
Consider a linear map $\kappa$:
\begin{equation}
   \kappa\;:\; {\cal T}\widehat{M} \;\to\; {\cal A}
\end{equation}
such that:
\begin{align}
   \mbox{im }\kappa \;& = \; {\cal A}_{[0,3]}
   \\   
   a\circ\kappa \;& = \;
   {\bf id}\;:\;{\cal T}\widehat{M}\rightarrow {\cal T}\widehat{M}
\end{align}
Notice that the following operator:
\begin{equation}\label{aPerp}
   a^{\perp} = {\bf id} - \kappa\circ a
\end{equation}
is the projection to $\mbox{ker}(a)$ along ${\cal A}_{[0,3]}$.

Let us unapply the anchor from the RHS of (\ref{PartialPlusZ}):
\begin{align}
   \widetilde{\bf L}_+ \;& 
   =  \partial_+ + 
   \widetilde{J}^{L[mn]}_{0+}t^0_{L[mn]} + 
   \widetilde{J}^{R[mn]}_{0+}t^0_{R[mn]} + 
   \widetilde{J}_+^{\alpha}{\bf D}^L_{\alpha} +
   \widetilde{\Pi}_+^m{\bf A}_m^L + \widetilde{\psi}_{\alpha+}{\bf W}^{\alpha}_L 
\end{align}
Notice that:
\begin{equation}
   ({\bf Q}_L + {\bf Q}_R)^2\widetilde{\bf L}_+  = 0
\end{equation}
Indeed, let us for example look at the $\lambda_L\lambda_L$ part: 
\begin{equation}
   {\bf Q}_L^2 = Q_L^2 + 
   {1\over 2}\lambda_L^{\alpha}\lambda_L^{\beta}
   \{{\bf D}^L_{\alpha}\;,\;{\bf D}^L_{\beta}\} = 
   Q_L^2 + {1\over 2}\lambda_L^{\alpha}\lambda_L^{\beta} R_{\alpha\beta}
\end{equation}
This implies that the calculation of the action of ${\bf Q}_L^2$ on $\widetilde{\bf L}_+$ does not lead 
out of ${\cal A}_{[0,3]}$. Therefore the calculation is the same as it would be under the
anchor, and the result is zero.

Also notice that:
\begin{equation}
   a\left(\left({\bf Q}_L + {\bf Q}_R\right)\widetilde{\bf L}_+\right) = 0
\end{equation}
However it is not true that $({\bf Q}_L + {\bf Q}_R)\widetilde{\bf L}_+ = 0$; we will therefore correct $\widetilde{\bf L}_+$ 
by adding to it some expression with zero anchor.

\subsection{Correction $\widetilde{\bf L}_+\to {\bf L}_+$}
\subsubsection{General theory}
Let us consider deforming:
\begin{equation}
   \widetilde{\bf L}_+\mapsto \widetilde{\bf L}_+ + \Delta{\bf L}_+
\end{equation}
where $\Delta{\bf L}_+$ does not contain the derivative $\partial_+$ and is an anchorless element
of ${\cal A}$ such that:
\begin{equation}
   ({\bf Q}_L + {\bf Q}_R)^2 \Delta{\bf L}_+ = 0
\end{equation}
We also require that $\Delta {\bf L}_+$ be $\bf h$-invariant. Let us denote ${\cal Y}_+$ the linear space 
of all expressions $\bf X_+\in {\cal A}$ satisfying the following properties:
\begin{enumerate}
\item $\bf X_+$ is $\bf h$-invariant
\item $\bf X_+$ has conformal dimension $(1,0)$
\item $({\bf Q}_L+ {\bf Q}_R)^2 {\bf X}_+ =0$
\end{enumerate} 
By definition $\Delta {\bf L}_+$ belongs to ${\cal Y}_+$.

\refstepcounter{Lemmas}
\label{LemmaCohomologyOnY}
\paragraph     {Lemma \arabic{Lemmas}:}
The cohomology of the operator ${\bf Q}_L + {\bf Q}_R$ acting in $\cal Y_+$ is zero.

\paragraph     {Proof:} 
Let us prove that the cohomology of ${\bf Q}_L$ is zero. First of all let us prove 
this statement in flat space. In flat space the algebroid is homogeneous, it
is defined by the same relations as $qU{\cal L}_{\rm tot}$. Given an expression annihilated
by ${\bf Q}_L$, let us consider the term with the lowest number of the letters $\nabla$.
It is $Q_L$-closed. Since the cohomology of $Q_L$ in the expressions of the 
conformal dimension $(1,0)$ is trivial, this means that this lowest order term
is exact. This completes the proof that the cohomology of ${\bf Q}_L$ is zero in flat
space.

In a general curved space, let us use the near-flat-space expansion 
(see \cite{Mikhailov:2012id} for details). For an element $\phi\in {\cal A}$ let us define its degree $\mbox{deg}(\phi)$ 
so that $\mbox{deg}(\theta)=\mbox{deg}(\lambda)=1$, $\mbox{deg}(w) =3$, $\mbox{deg}(x)=2$ and
$\mbox{deg}({\bf D}^L) = \mbox{deg}({\bf D}^R)=-1$. The proof follows from the following observations:
\begin{itemize}
\item $\mbox{deg}(({\bf Q}_L + {\bf Q}_R)\phi) \geq\mbox{deg}(\phi)$ 
\item the action in the associated graded space is the same as in flat space, 
\item we have just proven that the cohomology in flat space is zero. 
\end{itemize}
Lemma \ref{LemmaCohomologyOnY} implies the existence of such a ${\bf Y}_+\in {\cal Y}_+$ that:
\begin{equation}\label{DefYPlus}
   ({\bf Q}_L + {\bf Q}_R)(\widetilde{\bf L}_+ + {\bf Y}_+) = 0
\end{equation}
We therefore denote:
\begin{equation}
   {\bf L}_+ = \widetilde{\bf L}_+ + {\bf Y}_+
\end{equation}

\subsubsection{Explicit construction}
(If the reader is not familiar with the construction of the Lax operator for
$AdS_5\times S^5$ \cite{Berkovits:2000ph} we would recommend to first look at \cite{Mikhailov:2007mr}.)

\vspace{10pt}
\noindent
We will now show that the leading term of ${\bf Y}_+$ is in degree four, {\it i.e.}
${\bf L}_+$ is of the form:
\begin{align}
   {\bf L}_+ =\;& 
   \partial_+ + 
   J^{L[mn]}_{0+}t^0_{L[mn]} + 
   J^{R[mn]}_{0+}t^0_{R[mn]} + 
   \nonumber \\   
   & + \; J_+^{\alpha}{\bf D}^L_{\alpha} +
   \Pi_+^m{\bf A}_m^L + \psi_{\alpha+}{\bf W}^{\alpha}_L +
   \lambda_L^{\alpha}w^L_{\beta+}P_{\alpha\beta'}^{\alpha'\beta}
   \{\;{\bf D}^L_{\alpha'}\;,\;{\bf W}^{\beta'}_L\;\} 
   \label{LPlusIsOfDegreeFour}
\end{align}
where $P^{\alpha\beta}_{\gamma\delta}$ is the projector on the zero-form plus two-form. In other words, 
\begin{equation}\label{SolutionForYPlus}
   {\bf Y}_+ = \lambda_L^{\alpha}w^L_{\beta+}
   P_{\alpha\beta'}^{\alpha'\beta}a^{\perp}
   \{\;{\bf D}^L_{\alpha'}\;,\;{\bf W}^{\beta'}_L\;\} 
\end{equation}
where $a^{\perp}$ is defined in (\ref{aPerp}). We have to verify that ${\bf Y}_+$ satisfies (\ref{DefYPlus}). 
First of all, notice that $\psi_{\alpha+}$ in Eq. (\ref{LPlusIsOfDegreeFour}) is by its definition\footnote{We also use the fact that the projector $P^{\alpha'\beta}_{\alpha\beta'}$ only affects things in ${\cal A}_{[0,2]}$ --- see Section \ref{sec:BasicConsequences} } the same as 
$\psi_{\alpha+}$ defined in Eq. (\ref{PsiVsPsiTilde}). This implies that ${\bf Q}_L {\bf L}_+$ falls into ${\cal A}_{[0,3]}$. But at 
the same time, the anchor of ${\bf Q}_L {\bf L}_+$ is zero. This implies that ${\bf Q}_L{\bf L}_+=0$.

\refstepcounter{Lemmas}
\paragraph     {Lemma \arabic{Lemmas}:}
\begin{equation}
{\bf Q}_R{\bf L}_+ = 0
\end{equation}
\paragraph     {Proof}
Unfortunately we did not manage to prove it directly, but we have an indirect
argument.
Consider the action of ${\bf Q}_R$ on ${\bf L}_+$. Let us look at the leading term (which is 
in ${\cal A}_4$):
\begin{align}
   {\bf Q}_R \{\lambda_L^{\gamma}{\bf D}^L_{\gamma}\;,\;
   w^L_{\alpha +} {\bf W}_L^{\alpha}\} \;&=  
   \{\lambda_L^{\gamma}{\bf D}^L_{\gamma}\;,\;
   (Q_Rw^L_{\alpha +}) {\bf W}_L^{\alpha}\} \mbox{ mod } {\cal A}_{\leq 3}
\end{align}
The direct examination of the action shows that $Q_Rw^L_{\alpha+}=0$. (If $Q_Rw^L_{\alpha+}$ were
nonzero, the variation of the kinetic term $w_+\partial_-\lambda_L$ would result in the term 
with the structure $\lambda_R w^L_+\partial_-\lambda_L$ which would have nothing to cancel with.) 
Therefore ${\bf Q}_R{\bf L}_+$ falls into ${\cal A}_{[-1, 3]}$. Since the anchor is automatically zero, it
remains to prove that ${\bf Q}_R{\bf L}_+$ actually falls into ${\cal A}_{[0,3]}$. Let us look at the 
component of ${\bf Q}_R{\bf L}_+$ in grading $-1$. It is of the form:
\begin{equation}
   {\bf X}_+ = \phi_+^{\hat{\alpha}} {\bf D}^R_{\hat{\alpha}}
\end{equation}
We know that ${\bf Q}_L {\bf Q}_R {\bf L}_+ = 0$. This implies that $Q_L\phi_+^{\hat{\alpha}} = 0$. We conclude that
$\phi^{\hat{\alpha}}_+$ has conformal dimension $(1,0)$, ghost number $(0,1)$ and is $Q_L$-closed. But
there are not such operators, therefore $\phi_+^{\hat{\alpha}}=0$.

\subsubsection{Zero curvature}
\refstepcounter{Theorems}
\label{TheoremZeroCurv}
\paragraph     {Theorem \arabic{Theorems}:} 
\begin{equation}
   [{\bf L}_+,{\bf L}_-]=0
\end{equation}
Unfortunately we did not manage to prove it directly, but we have an indirect
argument. We know that $[{\bf L}_+,{\bf L}_-]$ is a dimension $(1,1)$ operator with components
in ${\cal A}_{[-4,4]}$, annihilated by both ${\bf Q}_L$ and ${\bf Q}_R$. Let us consider the highest 
component:
\begin{equation}
   X = [{\bf L}_+,{\bf L}_-]\mbox{ \tt\small mod } {\cal A}_{\leq 3}
\end{equation}
We know that $X$ is of the conformal dimension $(1,1)$ and ghost number zero. It 
follows that $Q_RX=0$. But there are no operators with such properties (as 
can be seen from the flat space limit). Therefore the components of $[{\bf L}_+,{\bf L}_-]$ 
span ${\cal A}_{-4,3}$. Then we can consider $[{\bf L}_+,{\bf L}_-]\mbox{ \tt\small mod } {\cal A}_{\leq 2}$ and so on.  

\subsection{Relation between integrated and unintegrated vertex}
There must be some analogue of the Koszul duality for algebroids, which should
imply that the cohomology of the BRST operator $\lambda_L^{\alpha}a({\bf D}^L_{\alpha}) + \lambda_R^{\hat{\alpha}}a({\bf D}^R_{\hat{\alpha}})$ is 
equivalent to the Lie algebroid cohomology of ${\cal A}$. Let us define ${\bf J}_+$ and ${\bf J}_-$ 
from the Lax pair:
\begin{equation}
   {\bf L}_{\pm} = {\partial\over\partial\tau^{\pm}} + {\bf J}_{\pm}
\end{equation}
Then, given a 2-cocycle $\psi$ representing the Lie algebroid cohomology, we can
construct the corresponding integrated vertex as in \cite{Chandia:2013kja}:
\begin{equation}
   U = \psi({\bf J}_+, {\bf J}_-)
\end{equation}
Moreover, the Koszul duality must also imply the consistency of the definition 
of the algebroid ${\cal A}$ (PBW). 

\vspace{7pt}

\noindent
We leave the details for future work.

\paragraph     {Is ${\cal A}$ an overkill?}
Notice that ${\bf J}_{\pm}$ only requires a small part of the 
${\cal A}$; indeed, ${\bf J}_+$ belongs to ${\cal A}_{[0,4]}$ and ${\bf J}_-$ belongs to ${\cal A}_{[-4,0]}$. This suggests that
our definition of ${\cal A}$ is quite an overkill.

\section*{Acknowledgements}
We are greatful to N.~Berkovits, R.~Heluani and M.~Movshev for explanations and 
useful/critical comments. The work was partially supported by the RFBR grant
15-01-99504 ``String theory and integrable systems''.

% \bibliographystyle{JHEP} \renewcommand{\refname}{Bibliography}
% \addcontentsline{toc}{section}{Bibliography}
% \bibliography{../andrei}

\def\cprime{$'$} \def\cprime{$'$}
\providecommand{\href}[2]{#2}\begingroup\raggedright\endgroup

\end{document}